\documentclass[aps,prx,twocolumn,groupedaddress]{revtex4-1}
\usepackage{dcolumn}
\usepackage{graphicx}
\usepackage{color}
\usepackage{amssymb}

\begin{document}


\title{Power-laws in the dynamic hysteresis of quantum nonlinear photonic resonators}


\author{W. Casteels$^1$, F. Storme$^1$, A. Le Boit\'e$^{1,2}$ and C. Ciuti$^1$}
\affiliation{$^1$Laboratoire Mat\'eriaux et Ph\'enom\`enes Quantiques, Universit\'e Paris Diderot-Paris-7 and CNRS, B\^atiment Condorcet, 10 rue Alice Domon et L\'eonie Duquet, 75205 Paris Cedex 13, France}
\affiliation{$^2$Institut f\"ur Theoretische Physik and IQST, Albert-Einstein Allee 11, Universit\"at Ulm, 89069 Ulm Germany}

\date{\today}

\begin{abstract}
We explore theoretically the physics of dynamic hysteresis for driven-dissipative nonlinear photonic resonators. In the regime where the semiclassical mean-field theory predicts bistability,  the exact steady-state density matrix is known to be unique, being a statistical mixture of two states: in particular, no static hysteresis cycle of the excited population occurs as a function of the driving intensity. Here, we predict that in the quantum regime a {\it dynamic} hysteresis with a rich phenomenology does appear when sweeping the driving amplitude in a finite time. The hysteresis area as a function of the sweep time reveals a double power-law decay, with a behavior qualitatively different from the mean-field predictions. The dynamic hysteresis power-law in the slow sweep limit defines a characteristic time, which depends dramatically on the size of the nonlinearity and on the frequency detuning between the driving and the resonator. In the strong nonlinearity regime, the characteristic time oscillates as a function of the intrinsic system parameters due to multiphotonic resonances. We show that the dynamic hysteresis for the considered class of driven-dissipative systems  is due to a non-adiabatic response region with connections to the Kibble-Zurek mechanism for quenched phase transitions. 
We also consider the case of two coupled driven-dissipative nonlinear resonators, showing that dynamic hysteresis and power-law behavior occur also in presence of correlations between resonators. Our theoretical predictions can be explored in a broad variety of physical systems, e.g., circuit QED superconducting resonators and semiconductor optical microcavities. 
\end{abstract}


\maketitle


\section{Introduction}
Since the first experimental realization \cite{PhysRevLett.36.1135}, optical bistability has been the subject of many investigations. One of the major theoretical breakthroughs was accomplished by Drummond and Walls, who provided an exact quantum solution for the steady-state density matrix of a single mode driven-dissipative nonlinear optical resonator \cite{0305-4470-13-2-034}. One of the main conclusions is somewhat surprising at first sight since it reveals a unique steady-state solution while the semiclassical mean-field theory \cite{semi_note} predicts a bistable regime with two stable branches. A large number of experimental studies in various systems have shown optical hysteresis cycles (to give just a few recent references, see, e.g., \cite{PhysRevLett.101.266402, PhysRevA.69.023809, PhysRevLett.113.057401, Paraiso:2010aa, PhysRevLett.93.207002, PhysRevLett.106.167002}) which seems in accordance with the  semiclassical mean-field approach. 

This apparent contradiction can be reconciled by realizing that fluctuations (quantum or classical) induce switching between the two branches resulting in a unique stationary mixed state solution \cite{PhysRevA.35.1729, PhysRevA.38.2409, PhysRevA.38.1349, PhysRevA.39.4675}. The switching is typically quantified by the lifetimes of the system in the different branches \cite{tunneling_note}. Without fluctuations, these lifetimes are infinite and produce the bistability at mean-field level. Note that the switching between the branches can be nicely visualized by considering individual quantum trajectories of the system \cite{Kerckhoff:11, :/content/aip/journal/apl/98/19/10.1063/1.3589994}. A similar behavior is obtained if classical fluctuations are considered such as for example thermal fluctuations \cite{:/content/aip/journal/rsi/80/11/10.1063/1.3224703} or a noisy drive \cite{PhysRevLett.113.057401, 2015arXiv150500932A}.  In general, it is  possible to define a transition point where the lifetimes of the two branches are equal. When the system is not at such transition point, one of the branches becomes increasingly more unstable than the other. These lifetimes can however become extremely large with respect to all other timescales resulting in quasi-bistability (the solutions are metastable) and explaining the success of the mean-field theory to describe the hysteresis. In recent years, photonic resonators with enhanced quantum nonlinearities have been developed \cite{RevModPhys.85.299}, in particular using superconducting quantum circuits or  semiconductor nanostructures as nonlinear media, paving the way to the experimental study of optical bistability in the quantum regime. 

In this paper, we report the surprising behavior of the hysteresis cycle in the quantum regime where the role played by quantum fluctuations and correlations becomes crucial. To explore such a physical problem, we have solved the time-dependent master equation for driven-dissipative nonlinear quantum resonators. By sweeping the drive amplitude in a finite time, we show that a dynamic hysteresis is obtained, even when the steady-state quantum solution is unique. Within a time-dependent mean-field approach, the area of the hysteresis cycle converges to the finite steady-state mean-field value for infinitely slow sweep with a deviation tending to zero as a power law, as it has been shown analytically and numerically for various classical models \cite{PhysRevLett.65.1873, doi:10.1137/S0036139995290733, PhysRevLett.74.2220}. Here we show that the behavior of the hysteresis area in the quantum regime is  qualitatively different in two ways: i) due to the uniqueness of the quantum steady-state solution, the area goes to zero in the limit of a very slow sweep; ii) the hysteresis area decays with increasing sweep time following a power law with an exponent that is different from the mean-field case for the same system parameters. We determine a characteristic time associated to the power-law decay and show its dramatic dependence on the size of the nonlinearity and frequency detuning of the driving. In the regime of strong photon nonlinearity, an oscillating behavior of the characteristic time as a function of the system parameters is shown to be due to the quantization of the photon field. We show that the power-law behavior of the dynamic hysteresis can be captured analytically by determining 
a non-adiabatic response region. 
Concerning the experimental implementations, the presented physics is expected to be applicable to a broad range of models of driven-dissipative nonlinear photonic models that exhibit bistability at the mean field level. Examples are: the driven-dissipative Jaynes-Cummings model \cite{7075, Carmichael2015}, an optomechanical cavity \cite{Dorsel1983, Meystre1985, Buchmann2012, Xu2015}, the driven-dissipative Dicke model \cite{Bowden1979, Klinder2015} and the micromaser \cite{Haroche}. Recently there has been a strong increase of interest in coupled nonlinear photonic modes arranged in large lattice structures for which the mean field approach typically predicts bistability \cite{PhysRevLett.103.033601, PhysRevLett.104.113601, PhysRevLett.110.163605, PhysRevLett.108.233603, PhysRevA.81.061801, RevModPhys.85.299, PhysRevLett.110.233601, PhysRevA.90.063821, Degenfeld2014, Houck:2012aa, PhysRevLett.108.206809, PhysRevA.87.053846, Mendoza2015}.  A better understanding of the role of quantum fluctuations is also crucial in the context of high speed optical switches of which the operation is based on optical bistability \cite{Notomi:05, Amo2010, Liew2010, Ballarini2013}.
 
\section{Theoretical framework and physical systems}  
Let us start by considering the quantum Hamiltonian ($\hbar = 1$) for a single-mode boson field with boson-boson interaction and coherent driving (treated within the rotating-wave approximation):
\begin{equation}
\hat{H}(t) = \omega_c \hat{a}^\dagger\hat{a}+\frac{U}{2}\hat{a}^\dagger\hat{a}^\dagger\hat{a}\hat{a} + F(t)e^{-i\omega_pt}\hat{a}^{\dagger} + F^*(t)e^{i\omega_pt}\hat{a}.
\label{Eq: SysHam}
\end{equation} 
This is also the Hamiltonian of a single-mode cavity field with a dispersive (Kerr) optical nonlinearity. Here, the boson operator $\hat{a}$ ($\hat{a}^{\dagger}$) annihilates (creates) an excitation in the cavity, while $\omega_c$ is the photon mode frequency, $U$ quantifies the photon-photon interaction, $F(t)$ is the time-dependent driving amplitude, and $\omega_p$ is the driving frequency. 

The dynamics with dissipation is described by the Lindblad master equation for the density matrix $\hat{\rho}(t)$:
\begin{eqnarray}
\frac{\partial\hat{\rho}(t)}{\partial t}= & i\left[\hat{\rho},\hat{H}(t)\right] + \frac{\gamma}{2}\left(1+n_{th} \right)\left( 2\hat{a}\hat{\rho}\hat{a}^\dagger - \hat{a}^\dagger\hat{a}\hat{\rho} - \hat{\rho}\hat{a}^\dagger\hat{a} \right) \nonumber\\
& + \frac{\gamma}{2}n_{th}\left( 2\hat{a}^\dagger\hat{\rho}\hat{a} - \hat{a}\hat{a}^\dagger\hat{\rho} - \hat{\rho}\hat{a}\hat{a}^\dagger \right),
\label{eq:Master}
\end{eqnarray}
where the term proportional to the commutator $\left[\hat{\rho},\hat{H}(t)\right]$ describes the quantum dynamics due to the Hamiltonian. The quantity $\gamma$ is the dissipation rate due to the coupling to the environment, $n_{th}$ the mean-number of thermal excitations at the resonator frequency $\omega_c$, namely 
$n_{th} = \left(e^{\beta\omega_c} - 1\right)^{-1},
$
with $\beta = (k_B T)^{-1}$, $T$ the bath temperature and $k_B$ the Boltzmann constant. From the density matrix the expectation value of an observable $\hat{O}$ can be calculated as $\langle \hat{O}\rangle = {\rm Tr}\left[ \hat{O}\hat{\rho}\right]$, where ${\rm Tr}$ denotes the trace. For the calculations, we will work in the frame rotating at the pump frequency for which the detuning $\Delta = \omega_p - \omega_c$ is the relevant parameter.   Notice that even in the rotating frame the Hamiltonian remains time-dependent due to time-dependence of the driving amplitude $F(t)$. 

For comparison, the mean-field equation for  the coherent field $\alpha (t) = \langle \hat{a} \rangle$ reads:
\begin{equation}
i\frac{\partial \alpha}{\partial t} = \left(\omega_c - i \frac{\gamma}{2} + U \left|\alpha \right|^2\right) \alpha + F(t) e^{-i\omega_pt}.
\label{MF}
\end{equation}
In the following, we will systematically compare the predictions of the exact solutions of the master equation (\ref{eq:Master}) to those obtained within the mean-field approximation in Eq. (\ref{MF}).

The present theoretical model is rather general and can be obtained, e.g.,  in a system consisting of a coherently driven linear cavity coupled to an ensemble of two-level atoms in the dispersive limit (large detuning between cavity and atom resonance frequencies with respect to the coupling strength) \cite{PhysRevA.23.2563}. Novel quantum optical systems with large nonlinearities such as superconducting quantum circuits and semiconductor microcavities have emerged in recent years \cite{RevModPhys.85.299}. In semiconductor pillars including quantum wells, a normalized $U/\gamma$ up to a few percent is within present capabilities. For these systems, a temperature $T=4$K (liquid helium bath) together with a cavity photon energy of $1.5$ eV give $\beta \omega_c \sim 10^4$ and a negligible number of thermal excitations: $n_{th} \approx 0$. In the context of circuit QED where a Josephson junction is used to introduce an effective nonlinearity \cite{PhysRevLett.93.207002, 2007cond.mat..2445B, :/content/aip/journal/rsi/80/11/10.1063/1.3224703, PhysRevLett.106.167002}, much larger nonlinearities can be be achieved with the possibility to reach the hardcore boson limit $\vert U \vert /\gamma \gg 1$ \cite{PhysRevLett.106.243601}. A typical dilution fridge temperature of $50$  mK  and a resonator frequency $\omega_c /(2\pi ) =  5$ GHz corresponds to $\beta\omega_c \simeq 4.8$ and $n_{th} \simeq 0.008$. Hence, thermal effects are in general small, but not completely negligible in circuit QED.

Note that here for simplicity we will not consider systems with absorptive optical nonlinearity and bistability (i.e., a cavity photon mode resonant with an electronic excitation like in the celebrated Jaynes-Cummings model) where quantum fluctuations also induce switching between the semiclassical branches \cite{7075, Carmichael2015}.

In order to solve numerically the master equation (\ref{eq:Master}), we have expressed the time-dependent density matrix in the basis of Fock number states $\vert n \rangle$, namely $\rho_{n,m}(t) = \langle n\left| \hat{\rho}(t) \right|m\rangle$. Convergence of the results have been carefully checked by increasing the cut-off number of photons. With this numerically exact integration method, we can typically explore regimes with number of photons up to few tens. 

\begin{figure}[t!]
  \includegraphics[scale=1.0]{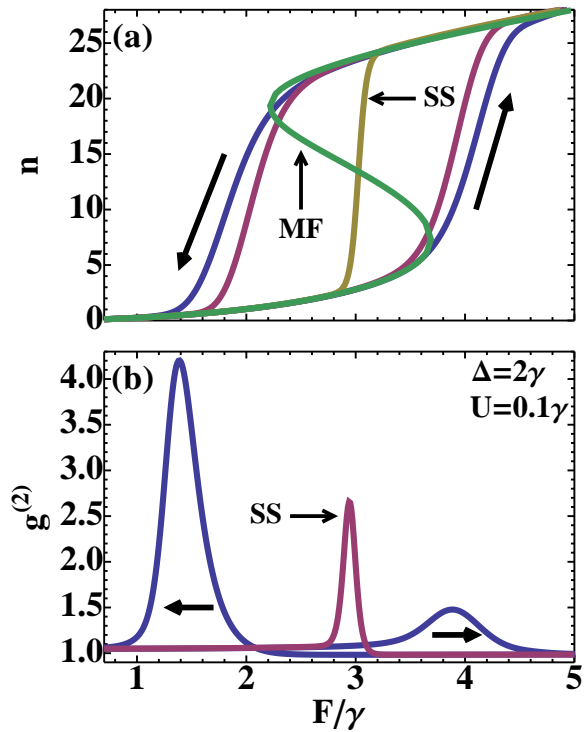}  
  \caption{\label{Fig1} (a) The photon population $n$  and (b) the $g^{(2)}$ second-order correlation function versus the driving amplitude $F$ (units of $\gamma$) for a single-mode driven-dissipative quantum resonator with a nonlinearity $U = 0.1\gamma$ and detuning $\Delta = 2\gamma$. In panel (a), the steady-state mean-field (MF) result and the quantum steady-state solution (SS) from Ref. \cite{0305-4470-13-2-034} are presented. The other two curves are dynamic hysteresis cycles predicted by the time-dependent quantum master equation obtained by using two different sweep times $t_s$ ($t_s/\Delta F = 10/\gamma^2$ for the curve with the largest hysteresis cycle and $t_s/\Delta F = 20 /\gamma^2$ for the smaller one). In panel (b) the steady-state solution is shown together with the result for a time-dependent sweep with $t_s/\Delta F = 10/\gamma^2$ (the arrows indicate the direction of the sweep).}
\end{figure}

\begin{figure}[t!]
  \includegraphics[scale=0.80]{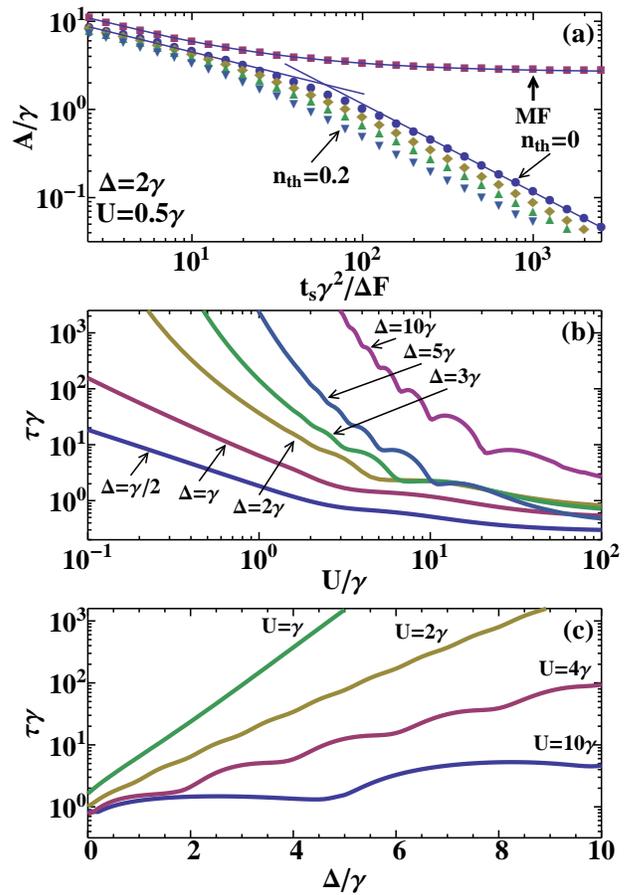}
  \caption{\label{Fig2} (a) The area $A$ of the hysteresis loop as a function of the sweep time $t_s$ (units of $\Delta F/\gamma^2$) for different temperatures (from bottom to top the thermal population $n_{th}$ is $0.2, 0.1, 0.05$ and $0$, corresponding respectively to $\beta \omega_c \simeq 1.8, 2.4, 3$ and $+ \infty$), together with the result from the mean-field approximation (MF) for $U=\gamma/2$ and $\Delta=2\gamma$. The full lines are power law fits to the different limiting regimes for which two separate power laws are observed. For large $t_s$ we find the behavior $A \propto t_s^{-1}$ while for small values of $t_s$ we find: $A \propto t_s^{-b}$ with a coefficient $b$ that depends on the system parameters. For the mean-field result we find an overall good agreement with $(A-A_0) \propto t_s^{-2/3}$ with $A_0 > 0$ the static hysteresis area. 
The characteristic timescale $\tau$, as determined from the behavior $A = (t_s/(\tau \Delta F))^{-1}$ for large $t_s$, is shown in (b) as a function of the nonlinearity $U$ (units of $\gamma$) for different values of the detuning $\Delta$ and in (c) as a function of the detuning $\Delta$ (units of $\gamma$) for different values of the nonlinearity $U$. Note the oscillating behavior with minima satisfying the $n$-photon resonance conditions: $Un(n-1)/2 = n\Delta $. The characteristic timescale $\tau$ in Fig. (a) is $115/\gamma$ . }
\end{figure}

\section{Numerical Results}

In order to study dynamical hysteresis phenomena, we consider a triangular modulation of the drive amplitude, namely  consisting of one sweep from $F_0$ to $F_0 + \Delta F$ and one from $F_0 + \Delta F$ back to $F_0$: 
\begin{equation}
F(t) = F_0 + \frac{t}{t_s}\Delta F \theta(t_s-t) - \frac{t - 2t_s}{t_s}\Delta F \theta(t - t_s),
\label{eq:Sweep}
\end{equation}
where $\theta(t)$ is the Heaviside step function and the time parameter $t_s $ is the sweep time. In practice the parameters $F_0$ and $\Delta F$ are always chosen such that the sweep covers the full range of the hysteresis. The master equation is solved in the time interval from $t = 0$ to $t = 2t_s$ with the steady-state solution at the pump intensity $F_0$ as an initial condition. Note that the presented results are in the zero temperature limit ($\beta\omega_c \rightarrow + \infty$ and $n_{th} \rightarrow 0$) unless explicitly stated otherwise.

\subsection{One resonator}
In Fig. \ref{Fig1} (a), we consider results for the excited population $n =\langle \hat{a}^{\dagger}\hat{a} \rangle$ using parameters for which the steady-state semiclassical mean-field (MF) solution exhibits the well-known bistability of the coherent population $\vert \alpha \vert^2$ as a function of the  driving amplitude (three branches with the middle one unstable). 
In stark contrast, the steady-state (SS) solution  \cite{0305-4470-13-2-034} of the master equation shows no hysteresis cycle for $n$.  
However, if we consider a time-dependent solution of the master equation with finite sweep time $t_s$, a clear {\it dynamic} hysteresis is found for $n(F(t))$.
The area of the dynamic hysteresis loop obtained with the solution of the master equation decreases for increasing $t_s$ (slower sweep). In the adiabatic limit of an infinitely slow sweep ($t_s \rightarrow + \infty$) the hysteresis disappears and the steady-state density-matrix solution of the master equation is recovered.
In Fig. \ref{Fig1} (b) the normalized second-order correlation function $g^{(2)} = \langle \hat{a}^{\dagger}\hat{a}^{\dagger}\hat{a}\hat{a} \rangle/n^2$ is presented as a function of the driving amplitude together with the steady-state result from Ref. \cite{0305-4470-13-2-034}. At a transition point, a sharp peak with $g^{(2)}$ significantly larger than $1$ occurs, a feature that cannot be captured by mean-field theory for which $g^{(2)} = 1$. For the time-dependent solution of the master equation, we find that this peak is shifted with respect to the steady-state solution to the region where the transition is seen in the density and it is more (less) pronounced for decreasing (increasing) $F$.

In order to study quantitatively the properties of dynamical hysteresis, we will evaluate the area $A$ of the hysteresis loop for a sweep with population $n_\uparrow(F)$ obtained for increasing driving amplitude $F$ and population $n_\downarrow(F)$ achieved for decreasing $F$:
\begin{equation}
A = \int_{F_0}^{F_0+\Delta F}dF \left| n_\downarrow(F)  - n_\uparrow(F)\right|.
\label{eq:Dist} 
\end{equation}

In Fig. \ref{Fig2} (a) the area $A$ is plotted as a function of the sweep time $t_s$ for different temperatures together with the result predicted by  the mean-field equation (\ref{MF}). Our results show that for relatively fast sweeps (small $t_s$) a reasonable agreement is found between the exact solution and the time-dependent mean-field result while for slower sweeps qualitatively different scaling laws are observed (the full lines in Fig. \ref{Fig2} (a) are power-law fitting curves). In the regime where the mean-field steady-state solutions exhibit bistability ($\Delta > \sqrt{3}/2\gamma$),  the time-dependent mean-field equation (\ref{MF}) predicts a dynamic hysteresis area which can be well fitted by the expression $(A - A_0)/\gamma \propto t_s^{-2/3}$, with $A_0$ the hysteresis area in the adiabatic limit ($t_s \rightarrow + \infty$). This is in agreement with the analytic result from Ref. \cite{PhysRevLett.65.1873} for a similar mean-field equation. In stark contrast, we find that the exact solution of the master equation has a double power-law behavior: for large $t_s$ the area scales as $A \propto t_s^{-1}$. In Section \ref{nonadiabatic}, we report an analytical derivation of the power law based on the determination of a non-adiabatic region. In  Appendix \ref{appA} , we present additional numerical results using a quasi-adiabatic approximation. In the regime where mean-field predicts no bistability ($\Delta < \sqrt{3}/2\gamma$) the dynamic mean-field result also exhibits the power law $A \propto t_s^{-1}$ for slow sweeps. Therefore, in contrast to the exact quantum solutions, we emphasize that the mean-field approach  predicts different scaling laws depending on the frequency detuning, a result in agreement with Refs. \cite{doi:10.1137/S0036139995290733, PhysRevLett.74.2220} treating mean-field models. Furthermore, we point out that our time-dependent quantum result exhibits another power law at small values of $t_s$  with a coefficient that depends on the system parameters (see Fig. \ref{Fig2} (a)). Note that this kind of double scaling law for the hysteresis area has also been observed in the context of dynamic transitions with magnetic materials \cite{RevModPhys.71.847, refId0}. Moreover, we see that the presence of a moderate thermal population (typical values of circuit QED experiments) gives the same power law behavior and just a moderate decrease of the hysteresis area as the temperature is increased. This is expected since thermal fluctuations also contribute to switching between the two branches.

The power-law behavior allows us to determine a characteristic timescale $\tau$ on which the hysteresis size will change significantly, namely through the expression $A = (t_s/(\tau \Delta F))^{-1}$, in the regime of large $t_s$ (see Fig. \ref{Fig2} (a)). For a sweep with $t_s \gamma/\Delta F \sim \tau$ the quantum fluctuations are expected to induce a significant deviation from the mean field result (as can also be seen in Fig. \ref{Fig2} (a) where $\tau/\gamma = 115$). The characteristic timescale $\tau$ is presented as a function of the nonlinearity for different values of the detuning in Fig. \ref{Fig2} (b) and as a function of the detuning for different values of the nonlinearity in Fig. \ref{Fig2} (c). We point out that the characteristic time $\tau$ can be orders of magnitude larger than the resonator lifetime $1/\gamma$ for small nonlinearities and/or large detuning. As a function of the detuning, an overall exponential increase of the characteristic hysteresis time is observed. Note that the dynamic hysteresis survives in the hard-core limit ($U \rightarrow \infty $) with $\tau$ converging to a finite value. Furthermore, a superimposed oscillating behavior of the characteristic time is predicted, which becomes more pronounced as the detuning or nonlinearity is increased. The minima correspond to system parameters satisfying the resonance condition $Un(n-1)/2 = n\Delta $, with $n$ a positive integer (giving the sequence $U = 2\Delta, \Delta, 2/3 \Delta,...$). These are the $n$-photon resonances \cite{PhysRevA.90.063821}, obtained when the energy of $n$ pump photons is equal to the energy of $n$ interacting photons in the resonator.

\begin{figure}[h]
  \includegraphics[scale=0.8]{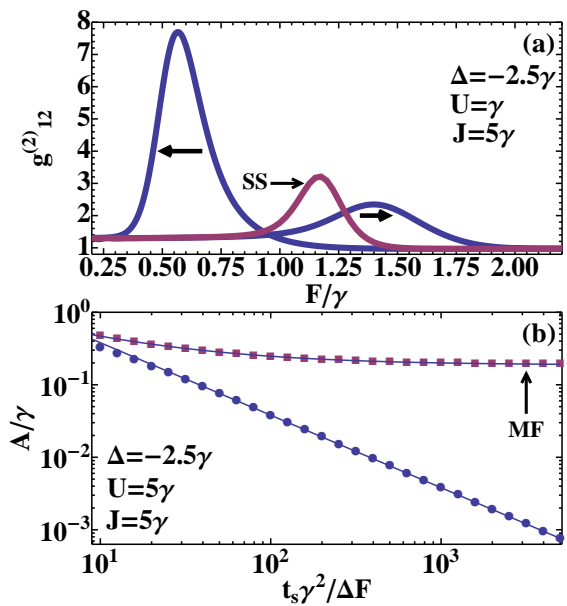}
  \caption{\label{Fig3} Results for a system consisting of two coupled identical resonators. (a) The inter-resonator correlation function $g^{(2)}_{12} $ as a function of the pump amplitude $F$ (in units of $\gamma$) during a sweep of the driving amplitude. Results are shown for the steady state (SS) and for a dynamic hysteresis with $t_s/(\Delta F) \gamma^2 = 30$ (the arrows indicate the direction of the sweep). (b) The hysteresis area $A$ associated to the photon population in one resonator as a function of the sweep time together with the mean-field (MF) result. The full lines are fits obtained with the following scaling laws: $A \propto t_s^{-1}$ for large $t_s$ and $(A - A_0)/\gamma \propto t_s^{-2/3}$ for the mean-field result. The other system parameters $\Delta$, $U$ and $J$ are specified in each panel.}
\end{figure}

\subsection{Two coupled resonators}
Now, we consider the case of two identical resonators coupled by the following hopping Hamiltonian:
\begin{equation}
\hat{H}_{hop} = -J\left(\hat{a}_1^\dagger\hat{a}_2 + h.c.\right),
\label{eq:Hop} 
\end{equation}
with $J$ the hopping parameter. Note that this model is currently experimentally realized, e.g., with semiconductor microcavities \cite{PhysRevLett.108.126403, PrivSaid} and with circuit QED resonators \cite{PhysRevLett.113.110502}. For this system, for sake of simplicity, we have considered the case when each resonator has the same driving term. Since adding a second resonator squares the dimension of the Hilbert space, in order to have exact results with arbitrary precision, we are restricted to lower photon numbers with respect to the single resonator case. A negative detuning $\Delta$ is considered which is blue-detuned with respect to the linear resonance corresponding to the 'bonding' single-particle state (having eigenfrequency $\omega = \omega_c - J$). As in the single-resonator case,  we find a dynamic transition around the regime where mean-field predicts bistability with the hysteresis cycle and a peak in the local second-order correlation function at the transition (qualitatively the same as observed for a single cavity in Fig. \ref{Fig1}). We also examined the inter-cavity normalized second-order correlation function: $g^{(2)}_{12} = \langle \hat{a}^{\dagger}_1\hat{a}^{\dagger}_2\hat{a}_2\hat{a}_1\rangle/(n_1 n_2)$, which is presented in Fig. \ref{Fig3} (a) for a temporal sweep and for the steady-state. Also for this non-local correlation function a peak is observed at the transition. In Fig. \ref{Fig3} (b) we have presented the hysteresis area, as defined in Eq. (\ref{eq:Dist}), where the population of one resonator mode is used (due to the symmetry the populations are equal in both resonators). This reveals qualitatively the same power laws as for a single cavity:  $A \propto t_s^{-1}$ for large $t_s$ and $(A - A_0) \propto t_s^{-2/3}$ for the dynamic mean-field result.
These results for two coupled cavities are relevant considering the recent  interest in strongly correlated photonic phases in arrays of cavities (see for example \cite{PhysRevLett.103.033601, PhysRevLett.104.113601, PhysRevLett.110.163605, PhysRevLett.108.233603, PhysRevA.81.061801, RevModPhys.85.299, PhysRevLett.110.233601, PhysRevA.90.063821, Houck:2012aa, PhysRevLett.108.206809, PhysRevA.87.053846}). It was shown that applying the Gutzwiller mean-field approach to a lattice of nonlinear driven-dissipative cavities predicts bistability \cite{PhysRevLett.110.233601, PhysRevA.90.063821}. Our exact results for the two-resonator system show that while the exact solution has no static hysteresis, a dynamic hysteresis does emerge in the quantum regime. A natural next step to be pursued in the future is a study of the dynamic hysteresis for larger arrays of coupled resonators. In particular, the dependence of the dynamic hysteresis on the inevitable presence of disorder, the effect of multimode dynamics and quasi-continuous spectrum of states is an open problem that would be interesting to explore.

\section{Analytical scaling behavior and connection with Kibble-Zurek mechanism\label{nonadiabatic}}
 In the previous section, we have presented a comprehensive set of numerical solutions of the master equation showing the rich properties of dynamic hysteresis for a driven-dissipative nonlinear quantum resonator.
In this section, we present an analytical demonstration of the power-law behavior and of the exponent.  Qualitatively, we show that the dynamical hysteresis is due to a non-adiabatic response of the considered system when the driving field is swept around the bistability region. 
Note that the approach presented here can be applied to a generic driven-dissipative system. 

When changing in time one parameter of an Hamiltonian system, by definition the response becomes non-adiabatic when the time scale of the change is much shorter than the time scale of the system internal dynamics. Such a time is proportional to the inverse of the energy gap between the ground state and the excited state manyfold. In the case of quantum phase transitions, the energy gap vanishes at the critical point (softening of the excitation mode) leading to a divergence of the corresponding internal dynamics time scale (critical slowing down). Therefore, when crossing a critical point, there is always a non-adiabatic response region around the transition. This property is at the heart of the celebrated Kibble-Zurek mechanism for the formation of topological defects in quenched quantum phase transitions \cite{Kibble1976, Zurek1985, Zurek2005} (see for example Ref. \cite{Dziarmaga2010} for a review). Recently, the Kibble-Zurek mechanism has been examined for the non-dissipative quantum Rabi model which consists of a zero-dimensional photonic mode coupled to a single two-level system \cite{Hwang2015}. 

\begin{figure}[h]
  \includegraphics[scale=0.55]{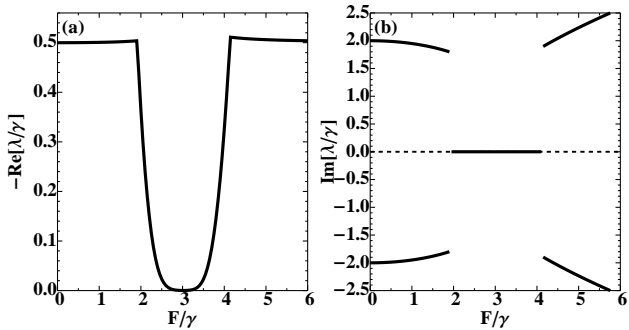}
  \caption{\label{Fig6} The real (a) and the imaginary (b) part of the Liouvilian eigenvalue $\lambda$ (in units of $\gamma$), corresponding, respectively, to the damping rate and the frequency of the excitation mode. In particular, we consider the least damped mode (different from the steady-state corresponding to $\lambda = 0$)
     as a function of the drive amplitude $F$ (in units of $\gamma$) for $U = 0.1\gamma$ and $\Delta = 2\gamma$. Around the transition point (at $F_c \approx 3\gamma$) the damping rate (real part) is strongly suppressed, while the imaginary part is exactly zero, indicating the presence of a soft diffusive mode. Away from the transition region there are two symmetric least damped modes with equal damping rates but opposite frequencies (the imaginary parts).}
\end{figure}

Since the class of systems we are studying in the present paper is of the driven-dissipative kind,  the gap of the Hamiltonian is not at all the relevant quantity. What is relevant here is the spectrum of the Liouvillian super-operator $\hat{L}$  associated to the master equation $\partial_t\hat{\rho} = \hat{L}\hat{\rho}$. We consider the eigenvalue equation for the Liouvillian superoperator:
\begin{equation}
 \hat{L} \hat{\rho}_{\lambda} = \lambda \hat{\rho}_{\lambda},
\end{equation}
where the eigenvalue $\lambda$ is in general complex. The steady-state density matrix corresponds to the eigenvalue $\lambda = 0$.  
Note that  the real part of $\lambda$ represents the damping of the excitation mode, while the imaginary part represents the oscillation frequency with respect to the steady-state.
Here, we will focus on the eigenvalue with the smallest non-zero real part since it is the least damped and determines the asymptotic relaxation to the steady-state. In Fig. \ref{Fig6} the real and imaginary part of this eigenvalue are presented as a function of the drive amplitude for $U = 0.1 \gamma$ and $\Delta = 2 \gamma$. 
Note that the the frequency gap (imaginary part) is $0$ in a finite interval around the value $F_c$ where the damping (real part) is also strongly suppressed, albeit reaching a finite minimum. This reveals that such a mode is soft and diffusive, i.e. because it is degenerate
in frequency with the steady-state and has a finite damping.  This kind of soft diffusive modes can appear in driven-dissipative systems  (in the case of  Bogoliubov excitations of driven-dissipative systems, see \cite{RevModPhys.85.299}). Hence, for a sweep of $F$ around $F_c$ the response of the system is expected to have a non-adiabatic contribution because of such frequency degeneracy. From this diffusive soft mode we determine the relaxation time $\tau_{R} = -1/Re[\lambda]$.
Note that the so-called tunneling time $\tau_T$ of bistability \cite{PhysRevA.35.1729, PhysRevA.38.2409, PhysRevA.38.1349, PhysRevA.39.4675,tunneling_note} corresponds to the maximal value of $\tau_{R}$, at the transition point.

\begin{figure}[h]
  \includegraphics[scale=0.7]{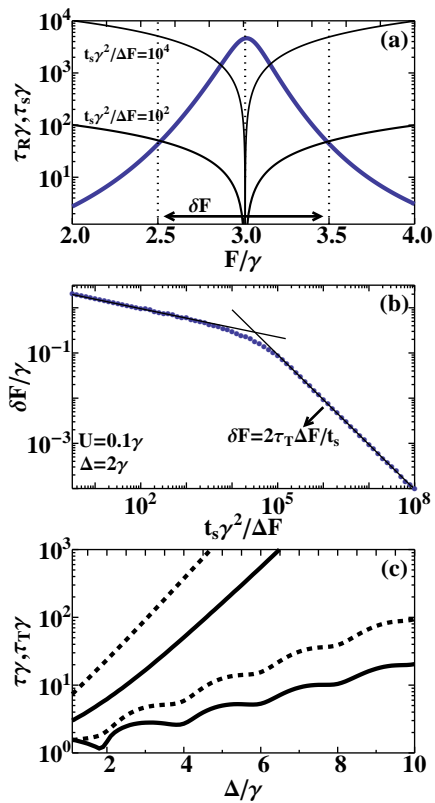}
  \caption{\label{Fig5} (a) The relaxation time $\tau_{R}$ (the curve peaked around $F_c \approx 3\gamma$)   is presented as a function of the drive amplitude together with the sweep timescale $\tau_s$ for two sweeps with different speeds for $U = 0.1\gamma$ and $\Delta = 2\gamma$. The non-adiabatic region with width $\delta F$ around $F_c$ is indicated for the  fastest sweep. (b) The width of the sweep $\delta F$ as a function of the sweep time together with the two power laws. (c) The tunneling time $\tau_T$ (full line) and the characteristic time $\tau$ (dashed line) versus the detuning $\Delta$ for nonlinearities $U/\gamma = 4$ (lower curves) and $U/\gamma = 1$ (upper curves).}
\end{figure}

For a time-dependent sweep of the drive amplitude,  we introduce the distance $\epsilon(t)$ from the value $F_c$, namely:
\begin{equation}
\epsilon(t) = F_c - F(t).
\end{equation}
We consider now a sweep of $F(t)$ linear in time from $F_c - \Delta F/2$ to $F_c + \Delta F/2$ with total time duration $t_s$. The normalized sweep rate reads \cite{Dziarmaga2010} \begin{equation}
\label{taus}
\left| \frac{\dot{\epsilon}(t)}{\epsilon(t)}\right| = \frac{{\Delta F}}{t_s} \frac{1}{\vert F_c - F(t) \vert}  = \frac{1}{\tau_s},
\end{equation}
which defines a sweep time scale $\tau_s(F) = t_s\vert F_c-F\vert/\Delta F$, which is plotted in solid thin line in Fig. \ref{Fig5} (a). 
Note that an alternative derivation of this expression can be obtained by equating the time from the transition point $F_c$ to the instantaneous relaxation time \cite{Zurek1985}.
Note that the sweep time scale $\tau_s$ differs from the duration $t_s$, because it contains also information on how much the driving amplitude is changed with respect to the transition point.

By generalizing the criterion used for the Kibble-Zurek mechanism in the context of equilibrium phase transitions, we expect that our driven-dissipative system enters a non-adiabatic regime when the sweep timescale $\tau_s$  becomes smaller than the relaxation time $\tau_{R}$. In Fig. \ref{Fig5} (a), we plot the relaxation time $\tau_{R}$  and the sweep timescale $\tau_s$ as a function of the driving amplitude $F$ for two different sweeps  ($t_s\gamma^2/\Delta F = 10^2$ and $10^4$). The corresponding non-adiabatic region of width $\delta F$ around $F_c$ is indicated in Fig. \ref{Fig5} (a) for the fastest sweep ($t_s\gamma^2/\Delta F = 10^2$). In Fig. \ref{Fig5} (b) the width $\delta F$ of the non-adiabatic region is presented versus the sweep time duration $t_s$, showing a double power-law, as also found for the area of the dynamic hysteresis. Furthermore, for slow sweeps the decay is proportional to $t_s^{-1}$, the same exponent as found before.
In the non-adiabatic region the system does not have the time to relax to the steady-state, resulting in an hysteretic behavior. The area of the hysteresis loop is therefore linked to the width $\delta F$ of the non-adiabatic region, as confirmed by our numerical results.

For $t_s \to + \infty$ (slow sweep limit), the non-adiabatic region size $\delta F \to 0$.
Hence, in this slow sweep limit, $\tau_{R} \to \tau_T$, the maximum of the  relaxation time. 
From Eq. (\ref{taus}) and imposing $\tau_s = \tau_{R} \simeq \tau_T$, we get the asymptotic expression for $\delta F$, namely:
\begin{equation}
\label{PowLAwKZ}
\delta F = 2\tau_{T} (t_s/\Delta F)^{-1} .
\end{equation} 
This formula is in excellent agreement with the results in Fig. \ref{Fig5} (b). We also note that for smaller values of $t_s$, this approximation fails and so a different power-law is expected, depending on the shape of the relaxation time vs $F$. 
This  shows that the exponent $-1$ of the slow sweep power law occurs because the {real part of the Liouvillian eigenvalue remains finite, but much smaller than all other characteristic frequencies of the system. The power law exponent is thus expected to change when also the real part of the Liouvillian eigenvalue vanishes; i.e. when $\tau_{T} \to \infty$, corresponding to a dissipative phase transition \cite{PhysRevA.86.012116}. This might occur in the thermodynamic limit of a large lattice of coupled driven-dissipative resonators.  
In Fig. \ref{Fig5} (c) the tunnelling time $\tau_T$ is compared with the characteristic time $\tau$ (see previous section and Fig. \ref{Fig2}) as a function of the detuning $\Delta$ for two values of the nonlinearity. This reveals qualitatively similar behavior with an overall exponential increase as a function of the detuning and oscillations due to the multi-photonic resonances.
 
For conservative systems with a finite energy gap the Kibble-Zurek mechanism breaks down for slow sweeps since the evolution becomes adiabatic \cite{Damski2005, Damski2006}.  In this case an effective description is provided by the Landau-Zener approximation for the evolution of a system through an avoided energy crossing \cite{Zener1932, Landau1932}. Applying the Kibble-Zurek mechanism results in a good agreement with the Landau-Zener result only for sufficiently fast sweeps \cite{Damski2005, Damski2006}. Note that the Landau-Zener formula for a dissipative excited state does not depend on the decay rate \cite{Akulin1992} and its applicability is connected to the existence of a finite gap for the frequency.

For the considered dissipative system on the other hand we find that the scaling laws based on a Kibble-Zurek-like approach for the non-adiabatic regime agree with the numerical results for the hysteresis area, also in the slow sweep limit. This shows that an adiabatic regime is never reached, no matter how slow is the sweep. At first sight this might seem in conflict with the results for the Kibble-Zurek mechanism for conservative systems since the real part of the Liouvillian gap remains finite. 
However, for dissipative systems it is the imaginary part of the Liouvillian eigenvalue that gives the excitation energy. 

\section{Conclusions}
We have investigated the time-dependent exact solutions of the quantum master equation for driven-dissipative nonlinear quantum resonators, thus including the role of quantum fluctuations and correlations. In particular, we have focussed on the regime where the semiclassical mean-field approximation predicts bistability and investigated temporal sweeps of the drive amplitude revealing dynamic hysteresis loops. {The hysteresis behavior, typically attributed to the semiclassical mean-field approach, was found to survive in the regime of small photon numbers and strong quantum fluctuations.} The time-dependent quantum solution, in contrast to predictions of mean-field approaches, shows that the hysteresis area as a function of the total sweep time has a double scaling law. These results have been shown to be robust with respect to thermal excitations for typical experimental temperatures. We have determined a characteristic time associated to the power-law decay of the dynamic hysteresis area, showing a rich behavior as a function of the nonlinearity and of the frequency detuning. We have  also considered two coupled driven-dissipative resonators, demonstrating that dynamic hysteresis and power-law decay occur also in presence of inter-cavity correlations. Importantly, we have demonstrated that the dynamic hysteresis is associated to a non-adiabatic response region with connections to the Kibble-Zurek mechanism for quenched phase transitions. We have been able to describe analytically the power-law behavior with scaling arguments and shown the role of a soft diffusive mode, i.e. having zero excitation energy, but a finite damping. This is a general picture, which  is expected  to apply to a broad class of driven-dissipative quantum systems. These exciting results can be a motivation to further investigate larger arrays of cavities at the dynamic transition. Given the emergence of several interesting systems with controllable quantum optical nonlinearities, the present predictions should stimulate exciting studies of dynamic hysteresis in the quantum regime.

We gratefully acknowledge discussions with A. Amo, J. Bloch, I. Carusotto, M. Hartmann, J. Lolli, S. Rodriguez, D. Sels and M. Wouters.
We  acknowledge  support  from  ERC  (via  the Consolidator  Grant ``CORPHO'' No.  616233),  from ANR (Project QUANDYDE No. ANR-11-BS10-0001), and from Institut Universitaire de France. A. L. B. also acknowledges support from  the SFB TRR/21 and the EU Integrating project SIQS.

\appendix
\section{Quasi-adiabatic approximation\label{appA}}
In this Appendix a quasi-adiabatic approximation is applied to study the dynamic hysteresis. From the Lindblad master equation (\ref{eq:Master}), the following equation of motion for the density $n$ can be derived:
\begin{equation}
\frac{\partial n}{\partial t} = -\gamma n  -i \left( F(t)\langle \hat{a}^\dagger \rangle - F^{*}(t)\langle \hat{a} \rangle \right).
\label{EOMn}
\end{equation}
As in the main text, we consider a linear sweep of the driving amplitude $F(t)$ (see Eq. (\ref{eq:Sweep})). For an adiabatic sweep ($t_s \rightarrow \infty$), the time-derivative tends to zero and all expectation values converge to the steady-state (SS) solution: 
\begin{eqnarray}
n_{t_s  \rightarrow \infty} & \rightarrow n_{SS}; \\
\langle \hat{a} \rangle _{t_s  \rightarrow \infty} & \rightarrow \langle \hat{a} \rangle_{SS}.
\end{eqnarray}
An analytical expression for the steady-state correlation functions was derived in Ref. \cite{0305-4470-13-2-034}. The result for the boson coherence $\langle a \rangle$ is (for sake of clarity we write the dependence on the driving amplitude $F$ explicitly):
\begin{equation}
\langle \hat{a} \rangle_{SS}(F)=  \frac{F}{\Delta + i \gamma/2} \frac{\mathcal{F}(1+c,c^*,8\left|\frac{F}{U}\right|^2)}{\mathcal{F}(1+c,c^*,8\left|\frac{F}{U}\right|^2)},
\label{SSa}
\end{equation}
with $c = -2(\Delta + i\gamma/2)/U$ and $\mathcal{F}(c,d,z)$ the hypergeometric function:
\begin{equation}
\mathcal{F}(c,d,z) = \sum_{n=0}^\infty \frac{\Gamma(c)\Gamma(d)}{\Gamma(c+n)\Gamma(d+n)}\frac{z^n}{n!},
\end{equation}
$\Gamma$ being the gamma special function. If the sweep is performed sufficiently slowly,  it is interesting to see what are the predictions of a quasi-adiabatic approximation. This corresponds to using the exact steady-state solution for $\langle \hat{a} \rangle$: $\langle \hat{a} \rangle \rightarrow \langle \hat{a} \rangle_{SS}(F(t))$, where the time-dependent driving amplitude $F(t)$ is used in Eq. (\ref{SSa}). By performing this substitution in the equation of motion for the density (\ref{EOMn}) and numerically solving the differential equation, the dynamic hysteresis can be examined within the quasi-adiabatic approximation. In Fig. \ref{Fig4} the resulting hysteresis area is compared to the exact numerical result for a set of parameters. The results clearly deviate, revealing that the quasi-adiabatic approximation does not capture the full dynamics. However, the same power law behavior $A \propto t_s^{-1}$ is found for sufficiently slow sweeps.  
\begin{figure}[h]
  \includegraphics[scale=0.8]{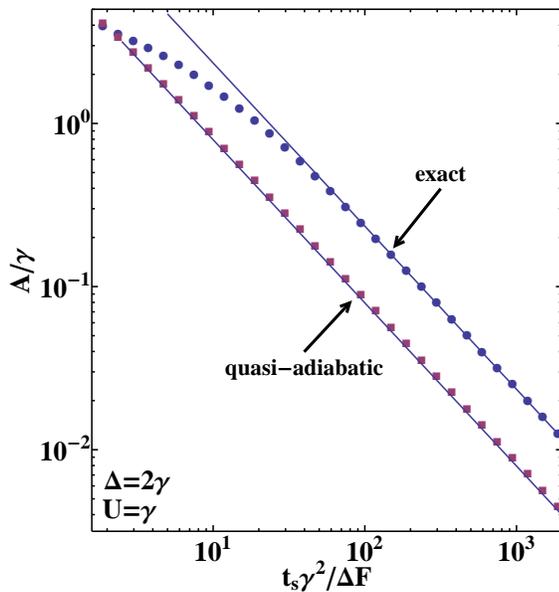}
  \caption{\label{Fig4} The hysteresis area $A$ as a function of the sweep time $t_s$ by solving exactly  the master equation (top curve) and by performing a quasi-adiabatic approximation (bottom curve) for $U = \gamma$ and $\Delta = 2\gamma$. The full lines are fits using the power law: $A \propto t_s^{-1}$ for large $t_s$.}
\end{figure}


\begin{thebibliography}{39}
\expandafter\ifx\csname natexlab\endcsname\relax\def\natexlab#1{#1}\fi
\expandafter\ifx\csname bibnamefont\endcsname\relax
  \def\bibnamefont#1{#1}\fi
\expandafter\ifx\csname bibfnamefont\endcsname\relax
  \def\bibfnamefont#1{#1}\fi
\expandafter\ifx\csname citenamefont\endcsname\relax
  \def\citenamefont#1{#1}\fi
\expandafter\ifx\csname url\endcsname\relax
  \def\url#1{\texttt{#1}}\fi
\expandafter\ifx\csname urlprefix\endcsname\relax\def\urlprefix{URL }\fi
\providecommand{\bibinfo}[2]{#2}
\providecommand{\eprint}[2][]{\url{#2}}

\bibitem[{\citenamefont{Gibbs et~al.}(1976)\citenamefont{Gibbs, McCall, and
  Venkatesan}}]{PhysRevLett.36.1135}
\bibinfo{author}{\bibfnamefont{H.~M.} \bibnamefont{Gibbs}},
  \bibinfo{author}{\bibfnamefont{S.~L.} \bibnamefont{McCall}},
  \bibnamefont{and} \bibinfo{author}{\bibfnamefont{T.~N.~C.}
  \bibnamefont{Venkatesan}}, \bibinfo{journal}{Phys. Rev. Lett.}
  \textbf{\bibinfo{volume}{36}}, \bibinfo{pages}{1135} (\bibinfo{year}{1976}),
  \urlprefix\url{http://link.aps.org/doi/10.1103/PhysRevLett.36.1135}.

\bibitem[{\citenamefont{Drummond and Walls}(1980)}]{0305-4470-13-2-034}
\bibinfo{author}{\bibfnamefont{P.~D.} \bibnamefont{Drummond}} \bibnamefont{and}
  \bibinfo{author}{\bibfnamefont{D.~F.} \bibnamefont{Walls}},
  \bibinfo{journal}{Journal of Physics A: Mathematical and General}
  \textbf{\bibinfo{volume}{13}}, \bibinfo{pages}{725} (\bibinfo{year}{1980}),
  \urlprefix\url{http://stacks.iop.org/0305-4470/13/i=2/a=034}.

\bibitem{semi_note} The semiclassical mean-field theory is obtained by replacing all the operators by $\mathbb{C}$-numbers. This corresponds to neglecting the quantum correlations and is typically a good approximation for large photon density.

\bibitem[{\citenamefont{Bajoni et~al.}(2008)\citenamefont{Bajoni, Semenova,
  Lema\^{i}tre, Bouchoule, Wertz, Senellart, Barbay, Kuszelewicz, and
  Bloch}}]{PhysRevLett.101.266402}
\bibinfo{author}{\bibfnamefont{D.}~\bibnamefont{Bajoni}},
  \bibinfo{author}{\bibfnamefont{E.}~\bibnamefont{Semenova}},
  \bibinfo{author}{\bibfnamefont{A.}~\bibnamefont{Lema\^{i}tre}},
  \bibinfo{author}{\bibfnamefont{S.}~\bibnamefont{Bouchoule}},
  \bibinfo{author}{\bibfnamefont{E.}~\bibnamefont{Wertz}},
  \bibinfo{author}{\bibfnamefont{P.}~\bibnamefont{Senellart}},
  \bibinfo{author}{\bibfnamefont{S.}~\bibnamefont{Barbay}},
  \bibinfo{author}{\bibfnamefont{R.}~\bibnamefont{Kuszelewicz}},
  \bibnamefont{and} \bibinfo{author}{\bibfnamefont{J.}~\bibnamefont{Bloch}},
  \bibinfo{journal}{Phys. Rev. Lett.} \textbf{\bibinfo{volume}{101}},
  \bibinfo{pages}{266402} (\bibinfo{year}{2008}),
  \urlprefix\url{http://link.aps.org/doi/10.1103/PhysRevLett.101.266402}.

\bibitem[{\citenamefont{Baas et~al.}(2004)\citenamefont{Baas, Karr, Eleuch, and
  Giacobino}}]{PhysRevA.69.023809}
\bibinfo{author}{\bibfnamefont{A.}~\bibnamefont{Baas}},
  \bibinfo{author}{\bibfnamefont{J.~P.} \bibnamefont{Karr}},
  \bibinfo{author}{\bibfnamefont{H.}~\bibnamefont{Eleuch}}, \bibnamefont{and}
  \bibinfo{author}{\bibfnamefont{E.}~\bibnamefont{Giacobino}},
  \bibinfo{journal}{Phys. Rev. A} \textbf{\bibinfo{volume}{69}},
  \bibinfo{pages}{023809} (\bibinfo{year}{2004}),
  \urlprefix\url{http://link.aps.org/doi/10.1103/PhysRevA.69.023809}.

\bibitem[{\citenamefont{Abbaspour et~al.}(2014)\citenamefont{Abbaspour,
  Trebaol, Morier-Genoud, Portella-Oberli, and
  Deveaud}}]{PhysRevLett.113.057401}
\bibinfo{author}{\bibfnamefont{H.}~\bibnamefont{Abbaspour}},
  \bibinfo{author}{\bibfnamefont{S.}~\bibnamefont{Trebaol}},
  \bibinfo{author}{\bibfnamefont{F.}~\bibnamefont{Morier-Genoud}},
  \bibinfo{author}{\bibfnamefont{M.~T.} \bibnamefont{Portella-Oberli}},
  \bibnamefont{and} \bibinfo{author}{\bibfnamefont{B.}~\bibnamefont{Deveaud}},
  \bibinfo{journal}{Phys. Rev. Lett.} \textbf{\bibinfo{volume}{113}},
  \bibinfo{pages}{057401} (\bibinfo{year}{2014}),
  \urlprefix\url{http://link.aps.org/doi/10.1103/PhysRevLett.113.057401}.

\bibitem[{\citenamefont{Para{\"\i}so et~al.}(2010)\citenamefont{Para{\"\i}so,
  Wouters, L{\'e}ger, Morier-Genoud, and Deveaud-Pl{\'e}dran}}]{Paraiso:2010aa}
\bibinfo{author}{\bibfnamefont{T.~K.} \bibnamefont{Para{\"\i}so}},
  \bibinfo{author}{\bibfnamefont{M.}~\bibnamefont{Wouters}},
  \bibinfo{author}{\bibfnamefont{Y.}~\bibnamefont{L{\'e}ger}},
  \bibinfo{author}{\bibfnamefont{F.}~\bibnamefont{Morier-Genoud}},
  \bibnamefont{and}
  \bibinfo{author}{\bibfnamefont{B.}~\bibnamefont{Deveaud-Pl{\'e}dran}},
  \bibinfo{journal}{Nat Mater} \textbf{\bibinfo{volume}{9}},
  \bibinfo{pages}{655} (\bibinfo{year}{2010}),
  \urlprefix\url{http://dx.doi.org/10.1038/nmat2787}.

\bibitem[{\citenamefont{Siddiqi et~al.}(2004)\citenamefont{Siddiqi, Vijay,
  Pierre, Wilson, Metcalfe, Rigetti, Frunzio, and
  Devoret}}]{PhysRevLett.93.207002}
\bibinfo{author}{\bibfnamefont{I.}~\bibnamefont{Siddiqi}},
  \bibinfo{author}{\bibfnamefont{R.}~\bibnamefont{Vijay}},
  \bibinfo{author}{\bibfnamefont{F.}~\bibnamefont{Pierre}},
  \bibinfo{author}{\bibfnamefont{C.~M.} \bibnamefont{Wilson}},
  \bibinfo{author}{\bibfnamefont{M.}~\bibnamefont{Metcalfe}},
  \bibinfo{author}{\bibfnamefont{C.}~\bibnamefont{Rigetti}},
  \bibinfo{author}{\bibfnamefont{L.}~\bibnamefont{Frunzio}}, \bibnamefont{and}
  \bibinfo{author}{\bibfnamefont{M.~H.} \bibnamefont{Devoret}},
  \bibinfo{journal}{Phys. Rev. Lett.} \textbf{\bibinfo{volume}{93}},
  \bibinfo{pages}{207002} (\bibinfo{year}{2004}),
  \urlprefix\url{http://link.aps.org/doi/10.1103/PhysRevLett.93.207002}.

\bibitem[{\citenamefont{Ong et~al.}(2011)\citenamefont{Ong, Boissonneault,
  Mallet, Palacios-Laloy, Dewes, Doherty, Blais, Bertet, Vion, and
  Esteve}}]{PhysRevLett.106.167002}
\bibinfo{author}{\bibfnamefont{F.~R.} \bibnamefont{Ong}},
  \bibinfo{author}{\bibfnamefont{M.}~\bibnamefont{Boissonneault}},
  \bibinfo{author}{\bibfnamefont{F.}~\bibnamefont{Mallet}},
  \bibinfo{author}{\bibfnamefont{A.}~\bibnamefont{Palacios-Laloy}},
  \bibinfo{author}{\bibfnamefont{A.}~\bibnamefont{Dewes}},
  \bibinfo{author}{\bibfnamefont{A.~C.} \bibnamefont{Doherty}},
  \bibinfo{author}{\bibfnamefont{A.}~\bibnamefont{Blais}},
  \bibinfo{author}{\bibfnamefont{P.}~\bibnamefont{Bertet}},
  \bibinfo{author}{\bibfnamefont{D.}~\bibnamefont{Vion}}, \bibnamefont{and}
  \bibinfo{author}{\bibfnamefont{D.}~\bibnamefont{Esteve}},
  \bibinfo{journal}{Phys. Rev. Lett.} \textbf{\bibinfo{volume}{106}},
  \bibinfo{pages}{167002} (\bibinfo{year}{2011}),
  \urlprefix\url{http://link.aps.org/doi/10.1103/PhysRevLett.106.167002}.

\bibitem[{\citenamefont{Risken et~al.}(1987)\citenamefont{Risken, Savage,
  Haake, and Walls}}]{PhysRevA.35.1729}
\bibinfo{author}{\bibfnamefont{H.}~\bibnamefont{Risken}},
  \bibinfo{author}{\bibfnamefont{C.}~\bibnamefont{Savage}},
  \bibinfo{author}{\bibfnamefont{F.}~\bibnamefont{Haake}}, \bibnamefont{and}
  \bibinfo{author}{\bibfnamefont{D.~F.} \bibnamefont{Walls}},
  \bibinfo{journal}{Phys. Rev. A} \textbf{\bibinfo{volume}{35}},
  \bibinfo{pages}{1729} (\bibinfo{year}{1987}),
  \urlprefix\url{http://link.aps.org/doi/10.1103/PhysRevA.35.1729}.

\bibitem[{\citenamefont{Vogel and Risken}(1988)}]{PhysRevA.38.2409}
\bibinfo{author}{\bibfnamefont{K.}~\bibnamefont{Vogel}} \bibnamefont{and}
  \bibinfo{author}{\bibfnamefont{H.}~\bibnamefont{Risken}},
  \bibinfo{journal}{Phys. Rev. A} \textbf{\bibinfo{volume}{38}},
  \bibinfo{pages}{2409} (\bibinfo{year}{1988}),
  \urlprefix\url{http://link.aps.org/doi/10.1103/PhysRevA.38.2409}.

\bibitem[{\citenamefont{Risken and Vogel}(1988)}]{PhysRevA.38.1349}
\bibinfo{author}{\bibfnamefont{H.}~\bibnamefont{Risken}} \bibnamefont{and}
  \bibinfo{author}{\bibfnamefont{K.}~\bibnamefont{Vogel}},
  \bibinfo{journal}{Phys. Rev. A} \textbf{\bibinfo{volume}{38}},
  \bibinfo{pages}{1349} (\bibinfo{year}{1988}),
  \urlprefix\url{http://link.aps.org/doi/10.1103/PhysRevA.38.1349}.

\bibitem[{\citenamefont{Vogel and Risken}(1989)}]{PhysRevA.39.4675}
\bibinfo{author}{\bibfnamefont{K.}~\bibnamefont{Vogel}} \bibnamefont{and}
  \bibinfo{author}{\bibfnamefont{H.}~\bibnamefont{Risken}},
  \bibinfo{journal}{Phys. Rev. A} \textbf{\bibinfo{volume}{39}},
  \bibinfo{pages}{4675} (\bibinfo{year}{1989}),
  \urlprefix\url{http://link.aps.org/doi/10.1103/PhysRevA.39.4675}.

\bibitem{tunneling_note} The quantum tunneling lifetimes introduced in Refs. \cite{PhysRevA.35.1729, PhysRevA.38.2409, PhysRevA.38.1349, PhysRevA.39.4675} are determined by calculating the lowest non-zero frequency eigenvalue of the Liouville super-operator of the Lindblad master equation. The Liouville super-operator $\hat{L}$ is defined via the master equation: $\partial_t\hat{\rho} = \hat{L}\hat{\rho}$, with $\hat{\rho}$ the density matrix. The quantum tunneling time is defined as the inverse of the lowest non-zero eigenvalue $\lambda$ that satisfies the equation $\hat{L}\hat{\rho}_\lambda = -\lambda \hat{\rho}_\lambda$, at the transition point. This approach is not able to describe dynamic hysteresis processes where the amplitude of the pump field is a time-dependent function $F(t)$ (even in the pump rotating frame).

\bibitem[{\citenamefont{Kerckhoff et~al.}(2011)\citenamefont{Kerckhoff, Armen,
  and Mabuchi}}]{Kerckhoff:11}
\bibinfo{author}{\bibfnamefont{J.}~\bibnamefont{Kerckhoff}},
  \bibinfo{author}{\bibfnamefont{M.~A.} \bibnamefont{Armen}}, \bibnamefont{and}
  \bibinfo{author}{\bibfnamefont{H.}~\bibnamefont{Mabuchi}},
  \bibinfo{journal}{Opt. Express} \textbf{\bibinfo{volume}{19}},
  \bibinfo{pages}{24468} (\bibinfo{year}{2011}),
  \urlprefix\url{http://www.opticsexpress.org/abstract.cfm?URI=oe-19-24-24468}.

\bibitem[{\citenamefont{Mabuchi}(2011)}]{:/content/aip/journal/apl/98/19/10.1063/1.3589994}
\bibinfo{author}{\bibfnamefont{H.}~\bibnamefont{Mabuchi}},
  \bibinfo{journal}{Applied Physics Letters} \textbf{\bibinfo{volume}{98}},
  \bibinfo{eid}{193109} (\bibinfo{year}{2011}),
  \urlprefix\url{http://scitation.aip.org/content/aip/journal/apl/98/19/10.1063/1.3589994}.

\bibitem[{\citenamefont{Vijay et~al.}(2009)\citenamefont{Vijay, Devoret, and
  Siddiqi}}]{:/content/aip/journal/rsi/80/11/10.1063/1.3224703}
\bibinfo{author}{\bibfnamefont{R.}~\bibnamefont{Vijay}},
  \bibinfo{author}{\bibfnamefont{M.~H.} \bibnamefont{Devoret}},
  \bibnamefont{and} \bibinfo{author}{\bibfnamefont{I.}~\bibnamefont{Siddiqi}},
  \bibinfo{journal}{Review of Scientific Instruments}
  \textbf{\bibinfo{volume}{80}}, \bibinfo{eid}{111101} (\bibinfo{year}{2009}),
  \urlprefix\url{http://scitation.aip.org/content/aip/journal/rsi/80/11/10.1063/1.3224703}.

\bibitem[{\citenamefont{{Abbaspour} et~al.}(2015)\citenamefont{{Abbaspour},
  {Sallen}, {Trebaol}, {Morier-Genoud}, {Portella-Oberli}, and
  {Deveaud}}}]{2015arXiv150500932A}
\bibinfo{author}{\bibfnamefont{H.}~\bibnamefont{{Abbaspour}}},
  \bibinfo{author}{\bibfnamefont{G.}~\bibnamefont{{Sallen}}},
  \bibinfo{author}{\bibfnamefont{S.}~\bibnamefont{{Trebaol}}},
  \bibinfo{author}{\bibfnamefont{F.}~\bibnamefont{{Morier-Genoud}}},
  \bibinfo{author}{\bibfnamefont{M.~T.} \bibnamefont{{Portella-Oberli}}},
  \bibnamefont{and}
  \bibinfo{author}{\bibfnamefont{B.}~\bibnamefont{{Deveaud}}},
  \bibinfo{journal}{ArXiv e-prints}  (\bibinfo{year}{2015}),
  \eprint{1505.00932}.

\bibitem[{\citenamefont{Carusotto and Ciuti}(2013)}]{RevModPhys.85.299}
\bibinfo{author}{\bibfnamefont{I.}~\bibnamefont{Carusotto}} \bibnamefont{and}
  \bibinfo{author}{\bibfnamefont{C.}~\bibnamefont{Ciuti}},
  \bibinfo{journal}{Rev. Mod. Phys.} \textbf{\bibinfo{volume}{85}},
  \bibinfo{pages}{299} (\bibinfo{year}{2013}),
  \urlprefix\url{http://link.aps.org/doi/10.1103/RevModPhys.85.299}.

\bibitem[{\citenamefont{Jung et~al.}(1990)\citenamefont{Jung, Gray, Roy, and
  Mandel}}]{PhysRevLett.65.1873}
\bibinfo{author}{\bibfnamefont{P.}~\bibnamefont{Jung}},
  \bibinfo{author}{\bibfnamefont{G.}~\bibnamefont{Gray}},
  \bibinfo{author}{\bibfnamefont{R.}~\bibnamefont{Roy}}, \bibnamefont{and}
  \bibinfo{author}{\bibfnamefont{P.}~\bibnamefont{Mandel}},
  \bibinfo{journal}{Phys. Rev. Lett.} \textbf{\bibinfo{volume}{65}},
  \bibinfo{pages}{1873} (\bibinfo{year}{1990}),
  \urlprefix\url{http://link.aps.org/doi/10.1103/PhysRevLett.65.1873}.

\bibitem[{\citenamefont{Broner et~al.}(1997)\citenamefont{Broner, Goldsztein,
  and Strogatz}}]{doi:10.1137/S0036139995290733}
\bibinfo{author}{\bibfnamefont{F.}~\bibnamefont{Broner}},
  \bibinfo{author}{\bibfnamefont{G.~H.} \bibnamefont{Goldsztein}},
  \bibnamefont{and} \bibinfo{author}{\bibfnamefont{S.~H.}
  \bibnamefont{Strogatz}}, \bibinfo{journal}{SIAM Journal on Applied
  Mathematics} \textbf{\bibinfo{volume}{57}}, \bibinfo{pages}{1163}
  (\bibinfo{year}{1997}), \eprint{http://dx.doi.org/10.1137/S0036139995290733},
  \urlprefix\url{http://dx.doi.org/10.1137/S0036139995290733}.

\bibitem[{\citenamefont{Hohl et~al.}(1995)\citenamefont{Hohl, van~der Linden,
  Roy, Goldsztein, Broner, and Strogatz}}]{PhysRevLett.74.2220}
\bibinfo{author}{\bibfnamefont{A.}~\bibnamefont{Hohl}},
  \bibinfo{author}{\bibfnamefont{H.~J.~C.} \bibnamefont{van~der Linden}},
  \bibinfo{author}{\bibfnamefont{R.}~\bibnamefont{Roy}},
  \bibinfo{author}{\bibfnamefont{G.}~\bibnamefont{Goldsztein}},
  \bibinfo{author}{\bibfnamefont{F.}~\bibnamefont{Broner}}, \bibnamefont{and}
  \bibinfo{author}{\bibfnamefont{S.~H.} \bibnamefont{Strogatz}},
  \bibinfo{journal}{Phys. Rev. Lett.} \textbf{\bibinfo{volume}{74}},
  \bibinfo{pages}{2220} (\bibinfo{year}{1995}),
  \urlprefix\url{http://link.aps.org/doi/10.1103/PhysRevLett.74.2220}.

\bibitem[{\citenamefont{Savage and Carmichael}(1988)}]{7075}
\bibinfo{author}{\bibfnamefont{C.}~\bibnamefont{Savage}} \bibnamefont{and}
  \bibinfo{author}{\bibfnamefont{H.~J.}~\bibnamefont{Carmichael}},
  \bibinfo{journal}{Quantum Electronics, IEEE Journal of}
  \textbf{\bibinfo{volume}{24}}, \bibinfo{pages}{1495} (\bibinfo{year}{1988}),
  ISSN \bibinfo{issn}{0018-9197}.


\bibitem[{\citenamefont{}(2015)}]{Carmichael2015}
\bibinfo{author}{\bibfnamefont{H.~J.}~\bibnamefont{Carmichael}}
  \bibinfo{journal}{Phys. Rev. X} \textbf{\bibinfo{volume}{5}},
  \bibinfo{pages}{031028} (\bibinfo{year}{2015}),
  \urlprefix\url{http://link.aps.org/doi/10.1103/PhysRevX.5.031028}.


\bibitem[{\citenamefont{Dorsel et~al.}(1983)\citenamefont{Dorsel, McCullen, Meystre, Vignes and Walther}}]{Dorsel1983}
\bibinfo{author}{\bibfnamefont{A.}~\bibnamefont{Dorsel}},
  \bibinfo{author}{\bibfnamefont{J. D.}~\bibnamefont{McCullen}},
  \bibinfo{author}{\bibfnamefont{P.}~\bibnamefont{Meystre}},
  \bibinfo{author}{\bibfnamefont{E.}~\bibnamefont{Vignes}} \bibnamefont{and}
  \bibinfo{author}{\bibfnamefont{H.}~\bibnamefont{Walther}},
  \bibinfo{journal}{Phys. Rev. Lett.} \textbf{\bibinfo{volume}{51}},
  \bibinfo{pages}{1550} (\bibinfo{year}{1983}),
  \urlprefix\url{http://link.aps.org/doi/10.1103/PhysRevLett.51.1550}.

\bibitem[{\citenamefont{Meystre et~al.}(1985)\citenamefont{Meystre, McCullen, Vignes and Wright}}]{Meystre1985}
\bibinfo{author}{\bibfnamefont{P.}~\bibnamefont{Meystre}},
  \bibinfo{author}{\bibfnamefont{J. D.}~\bibnamefont{McCullen}},
  \bibinfo{author}{\bibfnamefont{E.}~\bibnamefont{Vignes}} \bibnamefont{and}
  \bibinfo{author}{\bibfnamefont{E. M.}~\bibnamefont{Wright}},
  \bibinfo{journal}{J. Opt. Soc. Am. B} \textbf{\bibinfo{volume}{2}},
  \bibinfo{pages}{1830} (\bibinfo{year}{1985}),
  \urlprefix\url{https://www.osapublishing.org/josab/abstract.cfm?uri=josab-2-11-1830}.
  
\bibitem[{\citenamefont{Buchmann et~al.}(2012)\citenamefont{Buchmann, Zhang, Chiruvelli and Meystre}}]{Buchmann2012}
\bibinfo{author}{\bibfnamefont{L. F.}~\bibnamefont{Buchmann}},
  \bibinfo{author}{\bibfnamefont{L.}~\bibnamefont{Zhang}},
  \bibinfo{author}{\bibfnamefont{A.}~\bibnamefont{Chiruvelli}} \bibnamefont{and}
  \bibinfo{author}{\bibfnamefont{P.}~\bibnamefont{Meystre}},
  \bibinfo{journal}{Phys. Rev. Lett.} \textbf{\bibinfo{volume}{108}},
  \bibinfo{pages}{210403} (\bibinfo{year}{2012}),
  \urlprefix\url{http://link.aps.org/doi/10.1103/PhysRevLett.108.210403}.

\bibitem[{\citenamefont{Xu et~al.}(2015)citenamefont{Xu, Kemiktarak, Fan, Ragole, Lawall and Taylor}}]{Xu2015}
\bibinfo{author}{\bibfnamefont{H.}~\bibnamefont{{Xu}}},
  \bibinfo{author}{\bibfnamefont{U.} \bibnamefont{{Kemiktarak}}},
  \bibinfo{author}{\bibfnamefont{J.}~\bibnamefont{{Fan}}},
  \bibinfo{author}{\bibfnamefont{S.}~\bibnamefont{{Ragole}}},
  \bibinfo{author}{\bibfnamefont{J.}~\bibnamefont{{Lawall}}},
  \bibnamefont{and}
  \bibinfo{author}{\bibfnamefont{J. M.}~\bibnamefont{{Taylor}}},
  \bibinfo{journal}{eprint arXiv:1510.04971}  (\bibinfo{year}{2015}),
  \eprint{arXiv:1510.04971},
 \urlprefix\url{http://arxiv.org/abs/1510.04971}.

\bibitem[{\citenamefont{Bowden and Sung}(1995)\citenamefont{Bowden and Sung}}]{Bowden1979}
\bibinfo{author}{\bibfnamefont{C. M.} \bibnamefont{Bowden}}, \bibnamefont{and}
  \bibinfo{author}{\bibfnamefont{C. C.} \bibnamefont{Sung}},
  \bibinfo{journal}{Phys. Rev. A} \textbf{\bibinfo{volume}{19}},
  \bibinfo{pages}{2392} (\bibinfo{year}{1979}),
  \urlprefix\url{http://link.aps.org/doi/10.1103/PhysRevA.19.2392}.

\bibitem[{\citenamefont{Klinder et~al.}(1990)\citenamefont{Klinder, Keßler, Wolke, Mathey and Hemmerich}}]{Klinder2015}
\bibinfo{author}{\bibfnamefont{J.}~\bibnamefont{Klinder}},
  \bibinfo{author}{\bibfnamefont{H.}~\bibnamefont{Keßler}},
  \bibinfo{author}{\bibfnamefont{M.}~\bibnamefont{Wolke}}, 
  \bibinfo{author}{\bibfnamefont{L.}~\bibnamefont{Mathey}},\bibnamefont{and}
  \bibinfo{author}{\bibfnamefont{A.}~\bibnamefont{Hemmerich}},
  \bibinfo{journal}{PNAS} \textbf{\bibinfo{volume}{112}},
  \bibinfo{pages}{3290} (\bibinfo{year}{2015}),
  \urlprefix\url{http://www.pnas.org/content/112/11/3290}.

\bibitem[{\citenamefont{Haroche and Raimond}(2006)}]{Haroche}
\bibinfo{author}{\bibfnamefont{S.}~\bibnamefont{Haroche}} \bibnamefont{and}
  \bibinfo{author}{\bibfnamefont{J. M.}~\bibnamefont{Raimond}},
  \bibinfo{title}{\textit{Exploring the Quantum: Atoms, Cavities, and Photons}},
  \bibinfo{publisher}{Oxford Univ. Press} (\bibinfo{year}{2006}),
  \urlprefix\url{http://www.oxfordscholarship.com/view/10.1093/acprof:oso/9780198509141.001.0001/acprof-9780198509141}.


\bibitem[{\citenamefont{Carusotto et~al.}(2009)\citenamefont{Carusotto, Gerace,
  Tureci, De~Liberato, Ciuti, and Imamo\ifmmode~\check{g}\else
  \v{g}\fi{}lu}}]{PhysRevLett.103.033601}
\bibinfo{author}{\bibfnamefont{I.}~\bibnamefont{Carusotto}},
  \bibinfo{author}{\bibfnamefont{D.}~\bibnamefont{Gerace}},
  \bibinfo{author}{\bibfnamefont{H.~E.} \bibnamefont{Tureci}},
  \bibinfo{author}{\bibfnamefont{S.}~\bibnamefont{De~Liberato}},
  \bibinfo{author}{\bibfnamefont{C.}~\bibnamefont{Ciuti}}, \bibnamefont{and}
  \bibinfo{author}{\bibfnamefont{A.}~\bibnamefont{Imamo\ifmmode~\check{g}\else
  \v{g}\fi{}lu}}, \bibinfo{journal}{Phys. Rev. Lett.}
  \textbf{\bibinfo{volume}{103}}, \bibinfo{pages}{033601}
  (\bibinfo{year}{2009}),
  \urlprefix\url{http://link.aps.org/doi/10.1103/PhysRevLett.103.033601}.

\bibitem[{\citenamefont{Hartmann}(2010)}]{PhysRevLett.104.113601}
\bibinfo{author}{\bibfnamefont{M.~J.} \bibnamefont{Hartmann}},
  \bibinfo{journal}{Phys. Rev. Lett.} \textbf{\bibinfo{volume}{104}},
  \bibinfo{pages}{113601} (\bibinfo{year}{2010}),
  \urlprefix\url{http://link.aps.org/doi/10.1103/PhysRevLett.104.113601}.

\bibitem[{\citenamefont{Jin et~al.}(2013)\citenamefont{Jin, Rossini, Fazio,
  Leib, and Hartmann}}]{PhysRevLett.110.163605}
\bibinfo{author}{\bibfnamefont{J.}~\bibnamefont{Jin}},
  \bibinfo{author}{\bibfnamefont{D.}~\bibnamefont{Rossini}},
  \bibinfo{author}{\bibfnamefont{R.}~\bibnamefont{Fazio}},
  \bibinfo{author}{\bibfnamefont{M.}~\bibnamefont{Leib}}, \bibnamefont{and}
  \bibinfo{author}{\bibfnamefont{M.~J.} \bibnamefont{Hartmann}},
  \bibinfo{journal}{Phys. Rev. Lett.} \textbf{\bibinfo{volume}{110}},
  \bibinfo{pages}{163605} (\bibinfo{year}{2013}),
  \urlprefix\url{http://link.aps.org/doi/10.1103/PhysRevLett.110.163605}.

\bibitem[{\citenamefont{Nissen et~al.}(2012)\citenamefont{Nissen, Schmidt,
  Biondi, Blatter, T\"ureci, and Keeling}}]{PhysRevLett.108.233603}
\bibinfo{author}{\bibfnamefont{F.}~\bibnamefont{Nissen}},
  \bibinfo{author}{\bibfnamefont{S.}~\bibnamefont{Schmidt}},
  \bibinfo{author}{\bibfnamefont{M.}~\bibnamefont{Biondi}},
  \bibinfo{author}{\bibfnamefont{G.}~\bibnamefont{Blatter}},
  \bibinfo{author}{\bibfnamefont{H.~E.} \bibnamefont{T\"ureci}},
  \bibnamefont{and} \bibinfo{author}{\bibfnamefont{J.}~\bibnamefont{Keeling}},
  \bibinfo{journal}{Phys. Rev. Lett.} \textbf{\bibinfo{volume}{108}},
  \bibinfo{pages}{233603} (\bibinfo{year}{2012}),
  \urlprefix\url{http://link.aps.org/doi/10.1103/PhysRevLett.108.233603}.

\bibitem[{\citenamefont{Tomadin et~al.}(2010)\citenamefont{Tomadin,
  Giovannetti, Fazio, Gerace, Carusotto, T\"ureci, and
  Imamoglu}}]{PhysRevA.81.061801}
\bibinfo{author}{\bibfnamefont{A.}~\bibnamefont{Tomadin}},
  \bibinfo{author}{\bibfnamefont{V.}~\bibnamefont{Giovannetti}},
  \bibinfo{author}{\bibfnamefont{R.}~\bibnamefont{Fazio}},
  \bibinfo{author}{\bibfnamefont{D.}~\bibnamefont{Gerace}},
  \bibinfo{author}{\bibfnamefont{I.}~\bibnamefont{Carusotto}},
  \bibinfo{author}{\bibfnamefont{H.~E.} \bibnamefont{T\"ureci}},
  \bibnamefont{and} \bibinfo{author}{\bibfnamefont{A.}~\bibnamefont{Imamoglu}},
  \bibinfo{journal}{Phys. Rev. A} \textbf{\bibinfo{volume}{81}},
  \bibinfo{pages}{061801} (\bibinfo{year}{2010}),
  \urlprefix\url{http://link.aps.org/doi/10.1103/PhysRevA.81.061801}.

\bibitem[{\citenamefont{Le~Boit\'e et~al.}(2013)\citenamefont{Le~Boit\'e, Orso,
  and Ciuti}}]{PhysRevLett.110.233601}
\bibinfo{author}{\bibfnamefont{A.}~\bibnamefont{Le~Boit\'e}},
  \bibinfo{author}{\bibfnamefont{G.}~\bibnamefont{Orso}}, \bibnamefont{and}
  \bibinfo{author}{\bibfnamefont{C.}~\bibnamefont{Ciuti}},
  \bibinfo{journal}{Phys. Rev. Lett.} \textbf{\bibinfo{volume}{110}},
  \bibinfo{pages}{233601} (\bibinfo{year}{2013}),
  \urlprefix\url{http://link.aps.org/doi/10.1103/PhysRevLett.110.233601}.

\bibitem[{\citenamefont{Le~Boit\'e et~al.}(2014)\citenamefont{Le~Boit\'e, Orso,
  and Ciuti}}]{PhysRevA.90.063821}
\bibinfo{author}{\bibfnamefont{A.}~\bibnamefont{Le~Boit\'e}},
  \bibinfo{author}{\bibfnamefont{G.}~\bibnamefont{Orso}}, \bibnamefont{and}
  \bibinfo{author}{\bibfnamefont{C.}~\bibnamefont{Ciuti}},
  \bibinfo{journal}{Phys. Rev. A} \textbf{\bibinfo{volume}{90}},
  \bibinfo{pages}{063821} (\bibinfo{year}{2014}),
  \urlprefix\url{http://link.aps.org/doi/10.1103/PhysRevA.90.063821}.


\bibitem[{\citenamefont{Degenfeld-Schonburg and Hartmann}(2014)\citenamefont{Degenfeld-Schonburg, and Hartmann}}]{Degenfeld2014}
\bibinfo{author}{\bibfnamefont{P.}~\bibnamefont{Degenfeld-Schonburg}}  \bibnamefont{and}
  \bibinfo{author}{\bibfnamefont{M.~J.}~\bibnamefont{Hartmann}},
  \bibinfo{journal}{Phys. Rev. B} \textbf{\bibinfo{volume}{89}},
  \bibinfo{pages}{245108} (\bibinfo{year}{2014}),
  \urlprefix\url{http://link.aps.org/doi/10.1103/PhysRevB.89.245108}.


\bibitem[{\citenamefont{Houck et~al.}(2012)\citenamefont{Houck, Tureci, and
  Koch}}]{Houck:2012aa}
\bibinfo{author}{\bibfnamefont{A.~A.} \bibnamefont{Houck}},
  \bibinfo{author}{\bibfnamefont{H.~E.} \bibnamefont{Tureci}},
  \bibnamefont{and} \bibinfo{author}{\bibfnamefont{J.}~\bibnamefont{Koch}},
  \bibinfo{journal}{Nat Phys} \textbf{\bibinfo{volume}{8}},
  \bibinfo{pages}{292} (\bibinfo{year}{2012}),
  \urlprefix\url{http://dx.doi.org/10.1038/nphys2251}.

\bibitem[{\citenamefont{Umucal\ifmmode \imath \else~\i \fi{}lar and
  Carusotto}(2012)}]{PhysRevLett.108.206809}
\bibinfo{author}{\bibfnamefont{R.~O.} \bibnamefont{Umucal\ifmmode \imath
  \else~\i \fi{}lar}} \bibnamefont{and}
  \bibinfo{author}{\bibfnamefont{I.}~\bibnamefont{Carusotto}},
  \bibinfo{journal}{Phys. Rev. Lett.} \textbf{\bibinfo{volume}{108}},
  \bibinfo{pages}{206809} (\bibinfo{year}{2012}),
  \urlprefix\url{http://link.aps.org/doi/10.1103/PhysRevLett.108.206809}.

\bibitem[{\citenamefont{Grujic et~al.}(2013)\citenamefont{Grujic, Clark,
  Jaksch, and Angelakis}}]{PhysRevA.87.053846}
\bibinfo{author}{\bibfnamefont{T.}~\bibnamefont{Grujic}},
  \bibinfo{author}{\bibfnamefont{S.~R.} \bibnamefont{Clark}},
  \bibinfo{author}{\bibfnamefont{D.}~\bibnamefont{Jaksch}}, \bibnamefont{and}
  \bibinfo{author}{\bibfnamefont{D.~G.} \bibnamefont{Angelakis}},
  \bibinfo{journal}{Phys. Rev. A} \textbf{\bibinfo{volume}{87}},
  \bibinfo{pages}{053846} (\bibinfo{year}{2013}),
  \urlprefix\url{http://link.aps.org/doi/10.1103/PhysRevA.87.053846}.

\bibitem[{\citenamefont{Mendoza-Arenas et~al.}(2015)citenamefont{Mendoza-Arenas, Clark, Felicetti, Romero, Solano, Angelakis and Jaksch}}]{Mendoza2015}
\bibinfo{author}{\bibfnamefont{H.}~\bibnamefont{{Mendoza-Arenas}}},
  \bibinfo{author}{\bibfnamefont{U.} \bibnamefont{{Clark}}},
  \bibinfo{author}{\bibfnamefont{J.}~\bibnamefont{{Felicetti}}},
  \bibinfo{author}{\bibfnamefont{S.}~\bibnamefont{{Romero}}},
  \bibinfo{author}{\bibfnamefont{J.}~\bibnamefont{{Solano}}},
  \bibinfo{author}{\bibfnamefont{J.}~\bibnamefont{{Angelakis}}},
  \bibnamefont{and}
  \bibinfo{author}{\bibfnamefont{J. M.}~\bibnamefont{{Jaksch}}},
  \bibinfo{journal}{eprint arXiv:1510.06651}  (\bibinfo{year}{2015}),
  \eprint{arXiv:1510.06651},
 \urlprefix\url{http://arxiv.org/abs/1510.06651}.

\bibitem[{\citenamefont{Notomi et~al.}(2005)\citenamefont{Notomi, Shinya, Mitsugi, Kira, Kuramochi and Tanabe}}]{Notomi:05}
\bibinfo{author}{\bibfnamefont{M.}~\bibnamefont{Notomi}},
  \bibinfo{author}{\bibfnamefont{A.}~\bibnamefont{Shinya}},
  \bibinfo{author}{\bibfnamefont{S.}~\bibnamefont{Mitsugi}},
  \bibinfo{author}{\bibfnamefont{G.}~\bibnamefont{Kira}},
  \bibinfo{author}{\bibfnamefont{E.}~\bibnamefont{Kuramochi}} \bibnamefont{and}
  \bibinfo{author}{\bibfnamefont{T.}~\bibnamefont{Tanabe}},
  \bibinfo{journal}{Opt. Express} \textbf{\bibinfo{volume}{13}},
  \bibinfo{pages}{2678} (\bibinfo{year}{2005}),
  \urlprefix\url{https://www.osapublishing.org/oe/abstract.cfm?uri=oe-13-7-2678}.

\bibitem[{\citenamefont{Amo et~al.}(2010)\citenamefont{Amo, Liew, Adrados, Houdr\'e, Giacobino, Kavokin and Bramati}}]{Amo2010}
\bibinfo{author}{\bibfnamefont{A.}~\bibnamefont{Amo}},
  \bibinfo{author}{\bibfnamefont{T. C. H.}~\bibnamefont{Liew}},
  \bibinfo{author}{\bibfnamefont{C.}~\bibnamefont{Adrados}},
  \bibinfo{author}{\bibfnamefont{R.}~\bibnamefont{Houdr\'e}},
  \bibinfo{author}{\bibfnamefont{E.}~\bibnamefont{Giacobino}},
  \bibinfo{author}{\bibfnamefont{A. V..}~\bibnamefont{Kavokin}} \bibnamefont{and}
  \bibinfo{author}{\bibfnamefont{A.}~\bibnamefont{Bramati}},
  \bibinfo{journal}{Nature Photonics} \textbf{\bibinfo{volume}{4}},
  \bibinfo{pages}{361} (\bibinfo{year}{2010}),
  \urlprefix\url{http://www.nature.com/nphoton/journal/v4/n6/abs/nphoton.2010.79.html}.

\bibitem[{\citenamefont{Liew et~al.}(2010)\citenamefont{Liew, Kavokin, Ostatnick\'y, Kaliteevski, Shelykh and Abram}}]{Liew2010}
\bibinfo{author}{\bibfnamefont{T. C. H.}~\bibnamefont{Liew}},
  \bibinfo{author}{\bibfnamefont{A. V.}~\bibnamefont{Kavokin}},
  \bibinfo{author}{\bibfnamefont{T.}~\bibnamefont{Ostatnick\'y}},
  \bibinfo{author}{\bibfnamefont{M.}~\bibnamefont{Kaliteevski}},
  \bibinfo{author}{\bibfnamefont{I. A.}~\bibnamefont{Shelykh}} \bibnamefont{and}
  \bibinfo{author}{\bibfnamefont{R. A.}~\bibnamefont{Abram}},
  \bibinfo{journal}{Phys. Rev. B} \textbf{\bibinfo{volume}{82}},
  \bibinfo{pages}{033302} (\bibinfo{year}{2010}),
  \urlprefix\url{http://link.aps.org/doi/10.1103/PhysRevB.82.033302}.


\bibitem[{\citenamefont{Ballarini et~al.}(2013)\citenamefont{Ballarini, De Giorgi, Cancellieri, Houdr\'e, Giacobino, Cingoiani, Bramati, Gigli and Sanvitto}}]{Ballarini2013}
\bibinfo{author}{\bibfnamefont{D.}~\bibnamefont{Ballarini}},
  \bibinfo{author}{\bibfnamefont{M.}~\bibnamefont{De Giorgi}},
  \bibinfo{author}{\bibfnamefont{E.}~\bibnamefont{Cancellieri}},
  \bibinfo{author}{\bibfnamefont{R.}~\bibnamefont{Houdr\'e}},
  \bibinfo{author}{\bibfnamefont{E.}~\bibnamefont{Giacobino}},
  \bibinfo{author}{\bibfnamefont{R.}~\bibnamefont{Cingoiani}},
  \bibinfo{author}{\bibfnamefont{A.}~\bibnamefont{Bramati}},
  \bibinfo{author}{\bibfnamefont{G.}~\bibnamefont{Gigli}} \bibnamefont{and}
  \bibinfo{author}{\bibfnamefont{D.}~\bibnamefont{Sanvitto}},
  \bibinfo{journal}{Nature Communications} \textbf{\bibinfo{volume}{4}},
  \bibinfo{pages}{1778} (\bibinfo{year}{2013}),
  \urlprefix\url{http://www.nature.com/ncomms/journal/v4/n4/full/ncomms2734.html}.



\bibitem[{\citenamefont{Drummond and Walls}(1981)}]{PhysRevA.23.2563}
\bibinfo{author}{\bibfnamefont{P.~D.} \bibnamefont{Drummond}} \bibnamefont{and}
  \bibinfo{author}{\bibfnamefont{D.~F.} \bibnamefont{Walls}},
  \bibinfo{journal}{Phys. Rev. A} \textbf{\bibinfo{volume}{23}},
  \bibinfo{pages}{2563} (\bibinfo{year}{1981}),
  \urlprefix\url{http://link.aps.org/doi/10.1103/PhysRevA.23.2563}.

\bibitem[{\citenamefont{{Boaknin} et~al.}(2007)\citenamefont{{Boaknin},
  {Manucharyan}, {Fissette}, {Metcalfe}, {Frunzio}, {Vijay}, {Siddiqi},
  {Wallraff}, {Schoelkopf}, and {Devoret}}}]{2007cond.mat..2445B}
\bibinfo{author}{\bibfnamefont{E.}~\bibnamefont{{Boaknin}}},
  \bibinfo{author}{\bibfnamefont{V.~E.} \bibnamefont{{Manucharyan}}},
  \bibinfo{author}{\bibfnamefont{S.}~\bibnamefont{{Fissette}}},
  \bibinfo{author}{\bibfnamefont{M.}~\bibnamefont{{Metcalfe}}},
  \bibinfo{author}{\bibfnamefont{L.}~\bibnamefont{{Frunzio}}},
  \bibinfo{author}{\bibfnamefont{R.}~\bibnamefont{{Vijay}}},
  \bibinfo{author}{\bibfnamefont{I.}~\bibnamefont{{Siddiqi}}},
  \bibinfo{author}{\bibfnamefont{A.}~\bibnamefont{{Wallraff}}},
  \bibinfo{author}{\bibfnamefont{R.~J.} \bibnamefont{{Schoelkopf}}},
  \bibnamefont{and}
  \bibinfo{author}{\bibfnamefont{M.}~\bibnamefont{{Devoret}}},
  \bibinfo{journal}{eprint arXiv:cond-mat/0702445}  (\bibinfo{year}{2007}),
  \eprint{cond-mat/0702445}.

\bibitem[{\citenamefont{Lang et~al.}(2011)\citenamefont{Lang, Bozyigit,
  Eichler, Steffen, Fink, Abdumalikov, Baur, Filipp, da~Silva, Blais
  et~al.}}]{PhysRevLett.106.243601}
\bibinfo{author}{\bibfnamefont{C.}~\bibnamefont{Lang}},
  \bibinfo{author}{\bibfnamefont{D.}~\bibnamefont{Bozyigit}},
  \bibinfo{author}{\bibfnamefont{C.}~\bibnamefont{Eichler}},
  \bibinfo{author}{\bibfnamefont{L.}~\bibnamefont{Steffen}},
  \bibinfo{author}{\bibfnamefont{J.~M.} \bibnamefont{Fink}},
  \bibinfo{author}{\bibfnamefont{A.~A.} \bibnamefont{Abdumalikov}},
  \bibinfo{author}{\bibfnamefont{M.}~\bibnamefont{Baur}},
  \bibinfo{author}{\bibfnamefont{S.}~\bibnamefont{Filipp}},
  \bibinfo{author}{\bibfnamefont{M.~P.} \bibnamefont{da~Silva}},
  \bibinfo{author}{\bibfnamefont{A.}~\bibnamefont{Blais}},
  \bibnamefont{et~al.}, \bibinfo{journal}{Phys. Rev. Lett.}
  \textbf{\bibinfo{volume}{106}}, \bibinfo{pages}{243601}
  (\bibinfo{year}{2011}),
  \urlprefix\url{http://link.aps.org/doi/10.1103/PhysRevLett.106.243601}.


\bibitem[{\citenamefont{Chakrabarti and Acharyya}(1999)}]{RevModPhys.71.847}
\bibinfo{author}{\bibfnamefont{B.~K.} \bibnamefont{Chakrabarti}}
  \bibnamefont{and} \bibinfo{author}{\bibfnamefont{M.}~\bibnamefont{Acharyya}},
  \bibinfo{journal}{Rev. Mod. Phys.} \textbf{\bibinfo{volume}{71}},
  \bibinfo{pages}{847} (\bibinfo{year}{1999}),
  \urlprefix\url{http://link.aps.org/doi/10.1103/RevModPhys.71.847}.

\bibitem[{\citenamefont{{Huang, Zhigao} et~al.}(2005)\citenamefont{{Huang,
  Zhigao}, {Zhang, Fengming}, {Chen, Zhigao}, and {Du, Youwei}}}]{refId0}
\bibinfo{author}{\bibnamefont{{Huang, Zhigao}}},
  \bibinfo{author}{\bibnamefont{{Zhang, Fengming}}},
  \bibinfo{author}{\bibnamefont{{Chen, Zhigao}}}, \bibnamefont{and}
  \bibinfo{author}{\bibnamefont{{Du, Youwei}}}, \bibinfo{journal}{Eur. Phys. J.
  B} \textbf{\bibinfo{volume}{44}}, \bibinfo{pages}{423}
  (\bibinfo{year}{2005}),
  \urlprefix\url{http://dx.doi.org/10.1140/epjb/e2005-00141-4}.

\bibitem[{\citenamefont{Galbiati et~al.}(2012)\citenamefont{Galbiati, Ferrier,
  Solnyshkov, Tanese, Wertz, Amo, Abbarchi, Senellart, Sagnes, Lema\^{i}tre
  et~al.}}]{PhysRevLett.108.126403}
\bibinfo{author}{\bibfnamefont{M.}~\bibnamefont{Galbiati}},
  \bibinfo{author}{\bibfnamefont{L.}~\bibnamefont{Ferrier}},
  \bibinfo{author}{\bibfnamefont{D.~D.} \bibnamefont{Solnyshkov}},
  \bibinfo{author}{\bibfnamefont{D.}~\bibnamefont{Tanese}},
  \bibinfo{author}{\bibfnamefont{E.}~\bibnamefont{Wertz}},
  \bibinfo{author}{\bibfnamefont{A.}~\bibnamefont{Amo}},
  \bibinfo{author}{\bibfnamefont{M.}~\bibnamefont{Abbarchi}},
  \bibinfo{author}{\bibfnamefont{P.}~\bibnamefont{Senellart}},
  \bibinfo{author}{\bibfnamefont{I.}~\bibnamefont{Sagnes}},
  \bibinfo{author}{\bibfnamefont{A.}~\bibnamefont{Lema\^{i}tre}},
  \bibnamefont{et~al.}, \bibinfo{journal}{Phys. Rev. Lett.}
  \textbf{\bibinfo{volume}{108}}, \bibinfo{pages}{126403}
  (\bibinfo{year}{2012}),
  \urlprefix\url{http://link.aps.org/doi/10.1103/PhysRevLett.108.126403}.

\bibitem[{\citenamefont{Rodriguez et~al.}()\citenamefont{Rodriguez, Amo,
  Sagnes, Galopin, Lema\^itre, and Bloch}}]{PrivSaid}
\bibinfo{author}{\bibfnamefont{S.~R.~K.} \bibnamefont{Rodriguez}},
  \bibinfo{author}{\bibfnamefont{A.}~\bibnamefont{Amo}},
  \bibinfo{author}{\bibfnamefont{I.}~\bibnamefont{Sagnes}},
  \bibinfo{author}{\bibfnamefont{E.}~\bibnamefont{Galopin}},
  \bibinfo{author}{\bibfnamefont{A.}~\bibnamefont{Lema\^itre}},
  \bibnamefont{and} \bibinfo{author}{\bibfnamefont{J.}~\bibnamefont{Bloch}},
  \bibinfo{note}{private communications}.

\bibitem[{\citenamefont{Eichler et~al.}(2014)\citenamefont{Eichler, Salathe,
  Mlynek, Schmidt, and Wallraff}}]{PhysRevLett.113.110502}
\bibinfo{author}{\bibfnamefont{C.}~\bibnamefont{Eichler}},
  \bibinfo{author}{\bibfnamefont{Y.}~\bibnamefont{Salathe}},
  \bibinfo{author}{\bibfnamefont{J.}~\bibnamefont{Mlynek}},
  \bibinfo{author}{\bibfnamefont{S.}~\bibnamefont{Schmidt}}, \bibnamefont{and}
  \bibinfo{author}{\bibfnamefont{A.}~\bibnamefont{Wallraff}},
  \bibinfo{journal}{Phys. Rev. Lett.} \textbf{\bibinfo{volume}{113}},
  \bibinfo{pages}{110502} (\bibinfo{year}{2014}),
  \urlprefix\url{http://link.aps.org/doi/10.1103/PhysRevLett.113.110502}.


\bibitem[{\citenamefont{Kibble}(1976)\citenamefont{Kibble}}]{Kibble1976}
\bibinfo{author}{\bibfnamefont{T. W. B.}~\bibnamefont{Kibble}},
  \bibinfo{journal}{J. Phys. A} \textbf{\bibinfo{volume}{9}},
  \bibinfo{pages}{1387} (\bibinfo{year}{1976}),
  \urlprefix\url{http://iopscience.iop.org/article/10.1088/0305-4470/9/8/029/meta}.

\bibitem[{\citenamefont{Zurek}(1985)\citenamefont{Zurek}}]{Zurek1985}
\bibinfo{author}{\bibfnamefont{W. H.}~\bibnamefont{Zurek}},
  \bibinfo{journal}{Nature} \textbf{\bibinfo{volume}{317}},
  \bibinfo{pages}{505} (\bibinfo{year}{1985}),
  \urlprefix\url{http://www.nature.com/nature/journal/v317/n6037/abs/317505a0.html}.


\bibitem[{\citenamefont{Zurek et~al.}(2014)\citenamefont{Zurek, Dorner and Zoller}}]{Zurek2005}
\bibinfo{author}{\bibfnamefont{W. H.}~\bibnamefont{Zurek}},
  \bibinfo{author}{\bibfnamefont{U.}~\bibnamefont{Dorner}}, \bibnamefont{and}
  \bibinfo{author}{\bibfnamefont{P.}~\bibnamefont{Zoller}},
  \bibinfo{journal}{Phys. Rev. Lett.} \textbf{\bibinfo{volume}{95}},
  \bibinfo{pages}{105701} (\bibinfo{year}{2005}),
  \urlprefix\url{http://link.aps.org/doi/10.1103/PhysRevLett.95.105701}.

\bibitem[{\citenamefont{Dziarmaga}(2010)\citenamefont{Dziarmaga}}]{Dziarmaga2010}
\bibinfo{author}{\bibfnamefont{J.}~\bibnamefont{Dziarmaga}},
  \bibinfo{journal}{Advances in Physics} \textbf{\bibinfo{volume}{59}},
  \bibinfo{pages}{1063} (\bibinfo{year}{2010}),
  \urlprefix\url{http://www.tandfonline.com/doi/abs/10.1080/00018732.2010.514702}.

\bibitem[{\citenamefont{Hwang et~al.}(2014)\citenamefont{Hwang, Puebla and Plenio}}]{Hwang2015}
\bibinfo{author}{\bibfnamefont{M. J.}~\bibnamefont{Hwang}},
  \bibinfo{author}{\bibfnamefont{R.}~\bibnamefont{Puebla}}, \bibnamefont{and}
  \bibinfo{author}{\bibfnamefont{M. B.}~\bibnamefont{Plenio}},
  \bibinfo{journal}{Phys. Rev. Lett.} \textbf{\bibinfo{volume}{115}},
  \bibinfo{pages}{180404} (\bibinfo{year}{2015}),
  \urlprefix\url{http://link.aps.org/doi/10.1103/PhysRevLett.115.180404}.

\bibitem[{\citenamefont{Kessler et~al.}(2014)\citenamefont{Kessler, Giedke, Imamoglu, Yelin, Lukin and Cirac}}]{PhysRevA.86.012116}
\bibinfo{author}{\bibfnamefont{E. M.}~\bibnamefont{Kessler}},
\bibinfo{author}{\bibfnamefont{G.}~\bibnamefont{Giedke}},
\bibinfo{author}{\bibfnamefont{A.}~\bibnamefont{Imamoglu}},
\bibinfo{author}{\bibfnamefont{S. F.}~\bibnamefont{Yelin}},
  \bibinfo{author}{\bibfnamefont{M. D.}~\bibnamefont{Lukin}}, \bibnamefont{and}
  \bibinfo{author}{\bibfnamefont{J. I.}~\bibnamefont{Cirac}},
  \bibinfo{journal}{Phys. Rev. A} \textbf{\bibinfo{volume}{86}},
  \bibinfo{pages}{012116} (\bibinfo{year}{2012}),
  \urlprefix\url{http://link.aps.org/doi/10.1103/PhysRevA.86.012116}.


\bibitem[{\citenamefont{Damski}(2005)\citenamefont{Damski}}]{Damski2005}
\bibinfo{author}{\bibfnamefont{B.}~\bibnamefont{Damski}},
  \bibinfo{journal}{Phys. Rev. Lett.} \textbf{\bibinfo{volume}{95}},
  \bibinfo{pages}{035701} (\bibinfo{year}{2005}),
  \urlprefix\url{http://link.aps.org/doi/10.1103/PhysRevLett.95.035701}.


\bibitem[{\citenamefont{Damski et~al.}(2006)\citenamefont{Damski and Zurek}}]{Damski2006}
  \bibinfo{author}{\bibfnamefont{B.}~\bibnamefont{Damski}}, \bibnamefont{and}
  \bibinfo{author}{\bibfnamefont{W. H.}~\bibnamefont{Zurek}},
  \bibinfo{journal}{Phys. Rev. A} \textbf{\bibinfo{volume}{73}},
  \bibinfo{pages}{063405} (\bibinfo{year}{2006}),
  \urlprefix\url{http://link.aps.org/doi/10.1103/PhysRevA.73.063405}.

\bibitem[{\citenamefont{Zener}(1932)\citenamefont{Zener}}]{Zener1932}
\bibinfo{author}{\bibfnamefont{C.}~\bibnamefont{Zener}},
  \bibinfo{journal}{Proceedings of the Royal Society of London A: Mathematical, Physical and Engineering Sciences} \textbf{\bibinfo{volume}{137}},
  \bibinfo{pages}{696} (\bibinfo{year}{1932}),
  \urlprefix\url{http://rspa.royalsocietypublishing.org/royprsa/137/833/696.full.pdf}.

\bibitem[{\citenamefont{Landau}(1932)\citenamefont{Landau}}]{Landau1932}
\bibinfo{author}{\bibfnamefont{L.}~\bibnamefont{Landau}},
  \bibinfo{journal}{Physikalische Zeitschrift der Sowjetunion} \textbf{\bibinfo{volume}{2}},
  \bibinfo{pages}{46} (\bibinfo{year}{1932}).

\bibitem[{\citenamefont{Akulin et~al.}(2006)\citenamefont{Akulin and Schleich}}]{Akulin1992}
  \bibinfo{author}{\bibfnamefont{V. M.}~\bibnamefont{Akulin}}, \bibnamefont{and}
  \bibinfo{author}{\bibfnamefont{W. P.}~\bibnamefont{Schleich}},
  \bibinfo{journal}{Phys. Rev. A} \textbf{\bibinfo{volume}{46}},
  \bibinfo{pages}{4110} (\bibinfo{year}{1992}),
  \urlprefix\url{http://link.aps.org/doi/10.1103/PhysRevA.46.4110}.




\end{thebibliography}

\end{document}